\documentstyle[12pt]{article}

\newcommand{\lr}{{\mathcal{L}}}
\newcommand{\fr}{{\mathcal{F}}}
\newcommand{\hle}{\frac{\lambda}{2}}
\newcommand{\hl}{\lfloor\frac{\lambda}{2}\rfloor}
\newcommand{\hlo}{\frac{\lambda-1}{2}}
\renewcommand{\theequation}{\arabic{section}.\arabic{equation}}
\renewcommand{\(}{\begin{equation}}
\renewcommand{\)}{\end{equation}\vspace{-.05in}\linebreak}

\newcounter{saveeqn}
\newcounter{savealpheqn}
\newcommand{\alpheqn}{\setcounter{saveeqn}{\value{equation}}%
 \stepcounter{saveeqn}\setcounter{equation}{0}%
 \renewcommand{\theequation}{\mbox{\arabic{section}.\arabic{saveeqn}\alph{equation}}}
 \renewcommand{\)}{\end{equation}}}
\newcommand{\reseteqn}{\setcounter{equation}{\value{saveeqn}}%
 \renewcommand{\theequation}{\arabic{section}.\arabic{equation}}
 \renewcommand{\)}{\end{equation}\vspace{-.05in}\linebreak}
 \vspace{-.05in}\linebreak}
\newcommand{\aalpheqn}{\setcounter{saveeqn}{\value{equation}}%
 \stepcounter{saveeqn}\setcounter{equation}{0}%
 \renewcommand{\theequation}{\mbox{A.\arabic{saveeqn}\alph{equation}}}
  \renewcommand{\)}{\end{equation}}}
\newcommand{\areseteqn}{\setcounter{equation}{\value{saveeqn}}%
 \renewcommand{\theequation}{A.\arabic{equation}}
 \renewcommand{\)}{\end{equation}\vspace{-.05in}\linebreak}
 \vspace{-.05in}\linebreak}
\newcommand{\balpheqn}{\setcounter{saveeqn}{\value{equation}}%
 \stepcounter{saveeqn}\setcounter{equation}{0}%
 \renewcommand{\theequation}{\mbox{B.\arabic{saveeqn}\alph{equation}}}
  \renewcommand{\)}{\end{equation}}}
\newcommand{\breseteqn}{\setcounter{equation}{\value{saveeqn}}%
 \renewcommand{\theequation}{B.\arabic{equation}}
 \renewcommand{\)}{\end{equation}\vspace{-.05in}\linebreak}
 \vspace{-.05in}\linebreak}

\renewcommand{\+}{\hspace{-.03in}+\hspace{-.02in}}
\renewcommand{\thefootnote}{\alph{footnote}}

\mathchardef\endbar="375
\font\fivesans=cmss10 at 4.61pt
\font\sevensans=cmss10 at 6.81pt
\font\tensans=cmss10
\newfam\sansfam
\textfont\sansfam=\tensans\scriptfont\sansfam=\sevensans\scriptscriptfont
\sansfam=\fivesans
\def\sans{\fam\sansfam\tensans}
\def\Z{{\mathchoice
{\hbox{$\sans\textstyle Z\kern-0.4em Z$}}
{\hbox{$\sans\textstyle Z\kern-0.4em Z$}}
{\hbox{$\sans\scriptstyle Z\kern-0.3em Z$}}
{\hbox{$\sans\scriptscriptstyle Z\kern-0.2em Z$}}}}

\def\contr#1#2{\mathop{\vtop{\ialign{##\crcr
  $\hfil\displaystyle{#2}\hfil$\crcr\noalign{\kern3pt\nointerlineskip}
  \hspace{.09in}\rule[0in]{.01in}{.1in}\rule[0in]{#1in}{.01in}\rule[0in]{.01in}{.1in}\hskip6pt\crcr\noalign{\kern3pt}}}}}

\def\contrb#1#2#3{\mathop{\vtop{\ialign{##\crcr
  $\hfil\displaystyle{#3}\hfil$\crcr\noalign{\kern3pt\nointerlineskip}
  \hspace{#1in}\rule[0in]{.01in}{.1in}\rule[0in]{#2in}{.01in}\rule[0in]{.01in}{.1in}\hskip6pt\crcr\noalign{\kern3pt}}}}}

\def\namegroup#1{\begin{eqnarray}\label{#1}\nonumber\end{eqnarray}\vspace{-.5in}}

\def\rf#1{\ref{ref#1}}

\begin{document}
\begin{titlepage}
\begin{center}
April 14, 1999           \hfill UCB-PTH-99/17   \\
                                \hfill LBNL-43119    \\
                                \hfill hep-th/9904105    \\

\vskip .25in

{\large \bf General Virasoro Construction \\ on Orbifold Affine Algebra \\}

\vskip 0.3in
\def\thefootnote{\fnsymbol{footnote}}

J. EVSLIN, M. B. HALPERN, and J. E. WANG\footnote{E-Mail: hllywd2@physics.berkeley.edu}

\vskip 0.15in

{\em Department of Physics,
     University of California\\
     Berkeley, California 94720}\\
and\\
{\em Theoretical Physics Group\\
     Ernest Orlando Lawrence Berkeley National Laboratory\\
     University of California,
     Berkeley, California 94720}
        
\end{center}

\vskip .3in

\vfill

\begin{abstract}
We obtain the orbifold Virasoro master equation (OVME) at integer order $\lambda$, which summarizes the general Virasoro construction on orbifold affine algebra.  The OVME includes the Virasoro master equation when $\lambda=1$ and contains large classes of stress tensors of twisted sectors of conventional orbifolds at higher $\lambda$.  The generic construction is like a twisted sector of an orbifold (with non-zero ground state conformal weight) but new constructions are obtained for which we have so far found no conventional orbifold interpretation.\\ \\ \\
\end{abstract}

\vfill

\end{titlepage}
\setcounter{footnote}{0}
\renewcommand{\thefootnote}{\alph{footnote}}

%
%
%
%
%
%

\pagebreak
\renewcommand{\thepage}{\arabic{page}}

\section{Introduction}
The Virasoro master equation (VME) summarizes the general Virasoro construction$^{\rf{9},\rf{13}}$ on affine Lie algebra,
\begin{equation}
T(z)=L^{ab}:J_a(z)J_b(z): \hspace{.0666in}, \hspace{.3in} a,b=1,...,\textup{dim}g\ . \label{VMET}
\end{equation}
See Ref.~\rf{2} for a review of the VME and irrational conformal field theory.  Subsequent developments in this subject include semiclassical examples of conformal blocks$^{\rf{18}}$ in irrational conformal field theory, unification$^{\rf{19}-\rf{22}}$ with the general non-linear sigma model and the construction of the stress tensors of twisted sectors of cyclic permutation orbifolds$^{\rf{1}}$ from copies of the general Virasoro construction (\ref{VMET}).

Our starting point in this paper is the orbifold affine algebra$^{\rf{1}}$ at integer order $\lambda$, which was recently found in the twisted sectors of cyclic permutation orbifolds and includes affine Lie algebra when $\lambda=1$.  Following the suggestion of Ref.~\rf{1}, we consider here the general Virasoro construction on orbifold affine algebra
\begin{equation}
\hat{T}(z)=\sum_{r=0}^{\lambda-1}\lr_r^{ab}:\hat{J}_a^{(r)}(z)\hat{J}_b^{(-r)}(z):,\hspace{.2in}a,b=1,...,\textup{dim}g, \hspace{.2in} \lambda\in\Z^+
\end{equation}
where $\hat{J}_a^{(r)}(z)$, $r=0,...,\lambda -1$ are the orbifold currents.  When $\lambda=1$, the orbifold currents $\hat{J}_a^{(0)}(z)=J_a(z)$ satisfy affine Lie algebra and $\hat{T}(z)$ reduces to $T(z)$ in (\ref{VMET}).

The stress tensor $\hat{T}(z)$ is conformal when $\lr^{ab}_r$ satisfies the {\it{orbifold Virasoro master equation}} (OVME) at order $\lambda$, and the OVME contains the conformal field theories of the VME when $\lambda=1$.  At higher $\lambda$, the OVME contains large classes of stress tensors of twisted sectors of conventional permutation orbifolds, and many other constructions.  The generic construction of the OVME is apparently a twisted sector of an orbifold or a Ramond (spin) sector because the ground state conformal weights $\hat{\Delta}_0$ of the constructions
\begin{equation}
 L(m\geq 0)|0\rangle=\delta_{m,0}\hat{\Delta}_0|0\rangle,\hspace{.2in}\hat{\Delta}_0\neq 0
\end{equation}
are generically non-zero.  New unitary solutions of the OVME are obtained however, with irrational central charge and/or irrational conformal weights, for which we have so far found no conventional orbifold interpretation.

\section{Background}
\setcounter{equation}{0}
\subsection{Orbifold affine algebra}
We consider the \textit{orbifold affine algebra}$^{\rf{1}}$ $g_\lambda$ 
\namegroup{jalg}
\alpheqn
\begin{equation}
[\hat{J}^{(r)}_a(m\+\frac{r}{\lambda}),\hat{J}^{(s)}_b(n\+\frac{s}{\lambda})]\=if_{ab}^{ \ \ c}\hat{J}^{(r+s)}_c(m\+n\+\frac{r+s}{\lambda})\+\hat{G}_{ab}(m\+\frac{r}{\lambda})\delta_{m+n+\frac{r+s}{\lambda},0}
\end{equation}
\begin{equation}
m,n\in\Z,\hspace{.3in}a,b=1,...,\textup{dim}g,\hspace{.3in} r,s = 0,..,\lambda-1
\end{equation}
\begin{equation}
\hat{J}_a^{(r)}(m+\frac{r}{\lambda})|0\rangle=0\textrm{\ \ when\ \ }m+\frac{r}{\lambda}\geq 0
\end{equation}
\reseteqn
where the order $\lambda$ of $g_\lambda$ is any positive integer and $g_{\lambda=1}$ is affine Lie algebra$^{\rf{10}-\rf{6}}$.  The relations in (\ref{jalg}) are understood with the periodicity condition
\begin{equation}
\hat{J}_a^{(r\pm\lambda)}(m+\frac{r\pm\lambda}{\lambda})=\hat{J}_a^{(r)}(m\pm 1+\frac{r}{\lambda}) \label{percond}
\end{equation} 
and the state $|0\rangle$ is called the ground state of $g_\lambda$.  The quantities ${f_{ab}}^c$ are the structure constants of any semisimple Lie algebra $g=\oplus_Ig_I$, and the metric $\hat{G}_{ab}$ is
\begin{equation}
\hat{G}_{ab}=\oplus_I\hat{k}_I\eta_{ab}^I. \label{gdef}
\end{equation}
Here, $\eta^I_{ab}$ is the Killing metric of $g_I$ and $\hat{k}_I=\lambda k_I$ is the level of the associated orbifold affine subalgebra.  For applications we restrict ourselves to simple compact $g$,
\begin{equation}
\hat{G}_{ab}=\hat{k}\eta_{ab},\hspace{.2in}\hat{k}=\lambda k,\hspace{.2in}\hat{x}=\frac{2\hat{k}}{\psi^2}=\lambda x,\hspace{.2in}x=\frac{2k}{\psi^2} \label{xdef}
\end{equation}
where $\psi$ is the highest root of $g$ and unitarity$^{\rf{1}}$ requires that the invariant level $x$ be a positive integer.
 
We will also need the operator product form$^{\rf{1}}$ of $g_\lambda$
\alpheqn 
\begin{equation}
\hat{J}^{(r)}_a(z)\hat{J}^{(s)}_b(w)=
\frac{\hat{G}_{ab}\delta_{r+s,0\hspace{.05in}mod\hspace{.04in}\lambda}}{(z-w)^2}
+\frac{if_{ab}^{ \ \ c}\hat{J}_c^{(r+s)}(w)}{(z-w)} + O((z-w)^0) \label{JJOPE}
\end{equation}
\begin{equation}\hat{J}^{(r)}_a(z)=\sum_{m\in\Z}\hat{J}^{(r)}_a(m+\frac{r}{\lambda})z^{-1-m-r/\lambda}
\end{equation} \reseteqn
where $\hat{J}^{(r)}_a(z)$ are the local orbifold currents.

\subsection{Subalgebras of $g_\lambda$}
\label{sec-sub}
The algebra $g_\lambda$ contains a set of regularly embedded subalgebras $g_\eta\subset g_{\lambda}$,
\alpheqn
\begin{equation}
\hat{\hat{J\hspace{.045in}^{}}}\hspace{-.05in}^{(r)}_a(m+\frac{r}{\eta})=\hat{J}^{(\lambda r/\eta)}_a(m+\frac{\lambda r/\eta}{\lambda}),\hspace{.25in}r=0,...,\eta-1,\hspace{.25in}\eta\in{\Z}^+,\hspace{.25in}\frac{\lambda}{\eta}\in{\Z}^+
\end{equation}
\begin{eqnarray}
[\hat{\hat{J\hspace{.045in}^{}}}\hspace{-.05in}^{(r)}_a(m+\frac{r}{\eta}),\hat{\hat{J\hspace{.045in}^{}}}\hspace{-.05in}^{(s)}_b(n+\frac{s}{\eta})]
\hspace{-.02in} = \hspace{-.02in} i f_{ab}^{\ \ c} \hat{\hat{J\hspace{.045in}^{}}}\hspace{-.05in}^{(r+s)}_c(m\!+\!n\!+\!\frac{r+s}{\eta}) \!+\! \hat{G}_{ab}(m\!+\!\frac{r}{\eta})\delta_{m+n+\frac{r+s}{\eta},0}
\end{eqnarray}
\reseteqn
which is isomorphic to the set of order $\eta$ orbifold affine algebras, taken at levels $\{\hat{k}_I=\lambda k_I\}$.

The special case $\eta=1$ is known as the integral affine subalgebra$^{\rf{4},\rf{5},\rf{1}}$, whose generators 
\begin{equation}
\hat{\hat{J\hspace{.045in}^{}}}\hspace{-.05in}^{(0)}_a(m)=\hat{J}^{(0)}_a(m)
\end{equation}
form an affine Lie algebra at levels $\{\hat{k}_I\}$.
Related subalgebras include $h_{\eta}\subset g_{\lambda}$, $\lambda/\eta\in{\Z}^+$ where $h\subset g$ is any Lie subalgebra of $g$.  

There are many other subalgebras of $g_\lambda$, for example the algebra generated by
\begin{equation}
\hat{J}_a^{(\mu n\hspace{.05in}mod\hspace{.05in}\lambda)}(\lfloor\frac{\mu n}{\lambda}\rfloor+\frac{\mu n\hspace{.05in}mod\hspace{.05in}\lambda}{\lambda}),\hspace{.2in} n\in{\Z}, \hspace{.1in}\mu=0,1,2,... \hspace{.1in}
\end{equation}
where $\lfloor x \rfloor$ is the integer less than or equal to $x$.  When $\mu=\frac{\lambda}{\eta}\in\Z^+$, these are the generators 
\begin{equation}
\hat{\hat{J\hspace{.045in}^{}}}\hspace{-.05in}^{(r)}_a(m+\frac{r}{\eta})=\hat{J}^{(\mu r)}_a(m+\frac{\mu r}{\lambda}),\hspace{.5in}r=0,...,\eta-1
\end{equation}
of the $g_\eta$ subalgebras above.

\subsection{Normal ordering}

In Ref. \rf{1}, the (mother theory) normal-ordering convention
\begin{eqnarray}
:\hat{A}^{(r)}(z)\hat{B}^{(s)}(z):_M & \equiv & 
\sum_{m=-\infty}^{\infty}z^{-\Delta(A)-\Delta(B)-m-\frac{r+s}{\lambda}}
[\sum_{p\leq-\Delta(A)}\hat{A}^{(r)}(p+\frac{r}{\lambda})
\hat{B}^{(s)}(m-p+\frac{s}{\lambda})
\nonumber\\
& &+\sum_{p>-\Delta(A)}\hat{B}^{(s)}(m-p+\frac{s}{\lambda})\hat{A}^{(r)}(p+\frac{r}{\lambda})]
\end{eqnarray}
was used for the product of any two integer-moded orbifold principal primary fields, including the orbifold currents with (mother theory) conformal weights $\Delta(A)=\Delta(B)=1$.  We will use instead the OPE normal-ordering convention
\begin{equation}
:\hat{A}^{(r)}(z)\hat{B}^{(s)}(z): \hspace{.1in}\equiv
\ \oint_z\frac{(dx)}{x-z}\hat{A}^{(r)}(x)\hat{B}^{(s)}(z),\hspace{.5in} (dx)\equiv\frac{dx}{2\pi i}
\end{equation}
where the $x$ contour does not encircle the origin.  In the case of the orbifold currents, the two conventions are related by
\begin{equation}
:\hat{J}_a^{(r)}(z)\hat{J}_b^{(s)}(z): \hspace{.1in}\equiv \hspace{.1in}
:\hat{J}_a^{(r)}(z)\hat{J}_b^{(s)}(z):_M
-\frac{irf_{ab}^{\hspace{.11in}c}\hat{J}_c^{(r+s)}(z)}{\lambda z}
+\frac{r(\lambda-r)}{2\lambda^2z^2}\hat{G}_{ab}\delta_{r+s,0\hspace{.05in}mod\hspace{.04in}\lambda}
\end{equation}
and we note that the modes of the OPE normal-ordered bilinears satisfy
\alpheqn
\begin{equation}
:\hat{J}_a^{(r)}(z)\hat{J}_b^{(-r)}(z): \hspace{.1in}= \hspace{.1in} \sum_{m\in\Z} :\hat{J}_a^{(r)}\hat{J}_b^{(-r)}:(m) \ z^{-m-2}
\end{equation}
 \begin{equation}
:\hat{J}_a^{(r)}\hat{J}_b^{(-r)}:(m\geq 0)|0\rangle=\frac{r(\lambda-r)}{2\lambda^2}\hat{G}_{ab}\delta_{m,0}|0\rangle,\hspace{.3in}r=0,...,\lambda-1 \label{eq:delta}
\end{equation}
\reseteqn
because $:\hat{J}_a^{(r)}\hat{J}_b^{(-r)}:_M(m\geq 0)$ annihilates the ground state.

The current bilinears also have a symmetry (see App. A)
\begin{equation}:\hat{J}_{(a}^{(r)}(z)\hat{J}_{b)}^{(s)}(z): \hspace{.1in}=\hspace{.1in}:\hat{J}_{(a}^{(s)}(z)\hat{J}_{b)}^{(r)}(z): \label{keyident}
\end{equation}
which will play an important role in the general construction below.  (The symmetry holds for $:JJ:_M$ as well.)
\section{General Virasoro Construction}
\setcounter{equation}{0}
\subsection{Summary of the computation}
Following the suggestion of Ref. \rf{1}, we consider the candidate stress tensors
\namegroup{stresst}
\alpheqn
\begin{equation}
\hat{T}(z)=\sum_{r=0}^{\lambda-1}\lr_r^{ab}:\hat{J}_a^{(r)}(z)\hat{J}_b^{(-r)}(z): \hspace{.1in}
=\sum_{m\in{\Z}}L(m)z^{-m-2} \label{stress}
\end{equation}
\begin{equation}
\lr_r^{ab}=\lr_r^{ba},\hspace{.3in}r=0,...,\lambda-1 \label{orange}
\end{equation}
\reseteqn
where $\{\lr_r^{ab}\}$ is a set of $\lambda$ ``inverse inertia tensors''. 
Then, following Ref.$\hspace{.05in}$\rf{9},  we require that $\hat{T}$ satisfies the Virasoro algebra 
\begin{equation}
\hat{T}(z)\hat{T}(w)=\frac{\hat{c}/2}{(z-w)^4}+\frac{2\hat{T}(w)}{(z-w)^2}+\frac{\partial_w\hat{T}(w)}{(z-w)}+O((z-w)^0) \hspace{.1in}. \label{viralg} 
\end{equation}
This computation (see App. A) results in the restriction on $\lr_r^{ab}$
\namegroup{restric}
\alpheqn
\begin{equation}
\fr_r^{ab} =  \fr_r^{ac} \hat{G}_{cd} \fr_r^{db} - \frac{1}{2} \sum^{\lambda-1}_{s=0} \fr_s^{cd} [ \fr_{r+s}^{ef}f_{ce}^{\ \ a}f_{df}^{\ \ b} + f_{ce}^{\ \ f}f_{df}^{\ \ (a} \fr_r^{b) e}] \label{restrica}
\end{equation}
\begin{equation}
\fr_r^{ab} \equiv \lr_r^{ab}+\lr_{\lambda-r}^{ab}, \hspace{.3in} r=0,...,\lambda-1\label{restricb}
\end{equation}
\reseteqn
(where we have defined $\lr_{r \pm \lambda}^{ab}\equiv\lr_r^{ab}$) and the central charge,
\begin{equation}
\hat{c}=2 \hat{G}_{ab} \sum_{r=0}^{\lambda-1}\lr_r^{ab}. \label{cc}
\end{equation}
The form of the restriction (\ref{restric}), in terms of the combination $\fr_r^{ab}$, is a consequence of the symmetry (\ref{keyident}).  Any solution of the system (\ref{restric}) gives a conformal stress tensor $\hat{T}$, but this system is problematic because it has more unknowns than equations.

\subsection{Equivalent Solutions}

The system (\ref{restric}) determines only the combination $\fr_r^{ab}$ in (\ref{restricb}).  This means that (\ref{restric}) has a ``gauge invariance'' under the gauge transformation 
\namegroup{gaug}
\alpheqn
\begin{equation}
\lr_r^{ab}\rightarrow (\lr_r^{'})^{ab}=\lr_r^{ab}+\Lambda_r^{ab},\hspace{.3in} \lr_{\lambda-r}^{ab}\rightarrow (\lr_{\lambda-r}^{'})^{ab}=\lr_{\lambda-r}^{ab}-\Lambda_r^{ab} \hspace{.3in} \label{gauge}
\end{equation}
\begin{equation}
\fr_r^{ab} \rightarrow(\fr_r^{'})^{ab}= \fr_r^{ab}
\end{equation}
\begin{equation} r=1,...,\lfloor\hlo\rfloor
\end{equation}
\reseteqn
for any set of symmetric matrices $\Lambda_r^{ab}=\Lambda_r^{ba}$: If $\lr_r^{ab}$ is a solution of (\ref{restric}) then ${\lr_r^{'}}^{ab}$ is also a solution.
 
Using the symmetry (\ref{keyident}) again, we see that the gauge transformation (\ref{gaug}) is also an invariance of the stress tensor 
\[
\hat{T}=\left\{\begin{array}{cl}\lr_0^{ab}:\hat{J}_a^{(0)}\hat{J}_b^{(0)}: + \lr_{\lambda/2}^{ab}:\hat{J}_a^{(\lambda/2)}\hat{J}_b^{(\lambda/2)}: +  {\displaystyle\sum_{r=1}^{\frac{\lambda}{2}-1}} \fr_r^{ab}:\hat{J}_a^{(r)}\hat{J}_b^{(-r)}:\hspace{.1in} \textrm{for}\ \lambda\hspace{.08in}\textrm{even}\\
\lr_0^{ab}:\hat{J}_a^{(0)}\hat{J}_b^{(0)}: +  {\displaystyle\sum_{r=1}^{\frac{\lambda-1}{2}}} \fr_r^{ab} :\hat{J}_a^{(r)}\hat{J}_b^{(-r)}:\hspace{1.6in} \textrm{for}\ \lambda\hspace{.08in}\textrm{odd}
\end{array}\right.
\]
\vspace{-.4in}
\begin{equation}\end{equation}
and similarly for the central charge.  This tells us that all solutions in any given gauge orbit are physically equivalent, and we are entitled to choose a gauge.

\subsection{The OVME}
The most convenient gauge choice is 
\begin{equation} 
\lr_r^{ab} = \lr_{\lambda-r}^{ab}, \hspace{.1in} r=1,...,\lfloor\frac{\lambda-1}{2}\rfloor
\end{equation}
(choose $\Lambda_r^{ab}=(\lr_{\lambda-r}^{ab}-\lr_r^{ab})/2$) because this choice preserves the form of $\lr_r^{ab}$ given for the twisted sectors of cyclic orbifolds in Ref. \rf{1}. 

In this gauge we have 
\begin{equation}
\fr_r^{ab}=2\lr_r^{ab}
\end{equation}
and the Virasoro condition (\ref{viralg}) is summarized by the system
\namegroup{OVMEgrp}
\alpheqn
\begin{equation}
\hat{T}=\sum_{r=0}^{\lambda-1}\lr_r^{ab}:\hat{J}_a^{(r)}\hat{J}_b^{(-r)}: \label{stresstensor}\end{equation}
\begin{equation}
\lr_r^{ab} =  2 \lr_r^{ac}  \hat{G}_{cd} \lr_r^{db} - \sum^{\lambda-1}_{s=0} \lr_s^{cd} [\lr_{r+s}^{ef} f_{ce}^{\ \ a}f_{df}^{\ \ b} + f_{ce}^{\ \ f}f_{df}^{\ \ (a} \lr_r^{b) e}],\hspace{.15in} 0\leq r\leq\hl \label{OVME}
\end{equation}
\begin{equation}
\lr_r^{ab}=\lr_{r \pm \lambda}^{ab}=\lr_{\lambda \pm r}^{ab}, \hspace{.3in}\hl<r\leq\lambda\label{period}
\end{equation}
\begin{equation}
\hat{c}=2 \hat{G}_{ab} \sum_{r=0}^{\lambda-1}\lr_r^{ab}.
\end{equation}
\reseteqn
In what follows, we will refer to (\ref{OVME},c) as the {\textit{orbifold Virasoro master equation}} (OVME).  
 
As a check at $\lambda=1$, we define 
\begin{equation}
L^{ab}\equiv\lr^{ab}_0, \hspace{.3in} J_a\equiv\hat{J}_a^{(0)},\hspace{.3in}T\equiv\hat{T}
\end{equation}
and the OVME reduces to the Virasoro master equation$^{\rf{9},\rf{13}}$ (VME):
\namegroup{VME}
\alpheqn
\begin{equation}
T=L^{ab}:J_aJ_b: \label{VMEaa}
\end{equation}
\begin{equation}
L^{ab} =  2 L^{ac} G_{cd} L^{db} - L^{cd} L^{ef} f_{ce}^{\ \ a}f_{df}^{\ \ b} - L^{cd} f_{ce}^{\ \ f}f_{df}^{\ \ (a}L^{b) e} \label{VMEa}
\end{equation}
\begin{equation}
G_{ab}=\oplus_I k_I\eta_{ab}^I,\hspace{.2in} c=2G_{ab}L^{ab} \label{gabdef}
\end{equation}
\reseteqn
as it should since $g_{\lambda=1}$ is affine Lie algebra.

The solutions of the OVME are in one-to-one correspondence with the conformal stress tensors $\hat{T}$ in (\ref{stresstensor}), and we find
\begin{equation}
L(m\geq 0)|0\rangle=\delta_{m,0}\hat{\Delta}_0|0\rangle,\hspace{.2in}\hat{\Delta}_0 = \hat{G}_{ab} \sum_{r=0}^{\lambda -1} \lr_r^{ab} \frac{r (\lambda - r)}{2 \lambda^2}\label{delta}
\end{equation} 
where $\hat{\Delta}_0$ is the conformal weight of the ground state.  This conformal weight is generically nonzero so that the generic construction of the OVME is like a twisted sector of an orbifold or a Ramond spin sector.  The ground state conformal weight may occasionally vanish, e.g. the CFT's of the VME at $\lambda=1$ have $\hat{\Delta}_0=0$.

Using the constraints (\ref{period}) to pull back the $\lr$'s into the fundamental range\footnote{For $\lambda=1$ and $2$ the fundamental range in (\ref{frange}) is the original range in (\ref{orange}).},
\begin{equation}
\lr_r^{ab}, \hspace{.1in}r=0,...,\lfloor\frac{\lambda}{2}\rfloor \label{frange}
\end{equation}
the OVME has the explicit form:

\underline{For even $\lambda$}
\namegroup{ovmee}
\alpheqn
\begin{equation}
\hat{T} = \lr_0^{ab} :\hat{J}^{(0)}_a \hat{J}^{(0)}_b:  + \lr_{\lambda/2}^{ab} :\hat{J}_a^{(\lambda/2)} \hat{J}_b^{(\lambda /2)}:  + 2 \sum_{r=1}^{\frac{\lambda}{2}-1}\lr_r^{ab}:\hat{J}_a^{(r)}\hat{J}_b^{(-r)}:\end{equation}
\begin{eqnarray}
\lr_r^{ab}  &\hspace{-.15in}=\hspace{-.15in} &  2   \lr_r^{ac} \hat{G}_{cd} \lr_r^{db} - (2 \hspace{-.075in}\sum^{\hle-r-1}_{s=0} \hspace{-.05in} \lr_s^{cd} \lr_{r+s}^{ef}+\hspace{-.1in} \sum^{\hle}_{s=\hle-r} \hspace{-.05in} \lr_s^{cd} \lr_{\lambda-r-s}^{ef} + \sum^{r}_{s=0} \lr_s^{cd} \lr_{r-s}^{ef}-2\lr_0^{cd}\lr_r^{ef})f_{ce}^{\ \ a}f_{df}^{\ \ b} \nonumber\\ 
&& - (\lr_0^{cd} +  \lr_{\lambda/2}^{cd} + 2 \sum_{s=1}^{\frac{\lambda}{2}-1}\lr_s^{cd}  )f_{ce}^{\ \ f}f_{df}^{\ \ (a} \lr_r^{b) e},\hspace{.4in}r=0,...,\hle
\label{ovmee}\end{eqnarray}
\begin{equation}
\hat{c}=2   \hat{G}_{ab}\lr_0^{ab} + 2  \hat{G}_{ab} \lr_{\lambda/2}^{ab} + 4   \hat{G}_{ab}\sum_{r=1}^{\frac{\lambda}{2}-1}\lr_r^{ab}\end{equation}
\begin{equation}
\hat{\Delta}_0 =  \hat{G}_{ab} \lr_{\lambda/2}^{ab} + \hat{G}_{ab}  \sum_{r=1}^{\frac{\lambda}{2}-1} \lr_r^{ab} \frac{r (\lambda - r)}{\lambda^2} 
.\end{equation}
\reseteqn

\underline{For odd $\lambda$}
\namegroup{ovmeo}
\alpheqn
\begin{equation}
\hat{T} = \lr_0^{ab} :\hat{J}^{(0)}_a \hat{J}^{(0)}_b:   + 2 \sum_{r=1}^{\hlo}\lr_r^{ab}:\hat{J}_a^{(r)}\hat{J}_b^{(-r)}:
\end{equation}
\begin{eqnarray}
\lr_r^{ab}  & = &  2   \lr_r^{ac} \hat{G}_{cd} \lr_r^{db} - 
(2\sum^{\hlo-r}_{s=1} \lr_s^{cd} \lr_{r+s}^{ef}
+\sum^{\hlo}_{s=\hlo+1-r} \lr_s^{cd} \lr_{\lambda-r-s}^{ef}
+\sum^{r}_{s=0} \lr_s^{cd} \lr_{r-s}^{ef})f_{ce}^{\ \ a}f_{df}^{\ \ b}
\nonumber\\ & & - (\lr_0^{cd} + 2 \sum_{s=1}^{\hlo}\lr_s^{cd}  )f_{ce}^{\ \ f}f_{df}^{\ \ (a} \lr_r^{b) e},\hspace{.4in}r=0,...,\hlo \label{ovmeo}
\end{eqnarray}
\begin{equation}
\hat{c}=2  \hat{G}_{ab} \lr_0^{ab} + 4 \hat{G}_{ab}  \sum_{r=1}^{\frac{\lambda-1}{2}}\lr_r^{ab}
\end{equation}
\begin{equation}
\hat{\Delta}_0 =  \hat{G}_{ab} \sum_{r=1}^{\frac{\lambda-1}{2}} \lr_r^{ab}  \frac{r (\lambda - r)}{\lambda^2} 
.\end{equation}
\reseteqn
In these relations, we have used the convention that a sum is zero when its lower limit is greater than its upper limit.  The forms (\ref{ovmee}) and (\ref{ovmeo}) of the OVME are useful for computational purposes, while the form (\ref{OVME}) is useful for studying general properties of the system.

\section{Properties of the OVME}
\setcounter{equation}{0}

\subsection{Counting}

The OVME is a set of
\begin{equation}
n(g,\lambda)=(\lfloor \frac{\lambda}{2} \rfloor +1)\frac{\textup{dim}g (\textup{dim}g +1)}{2} \label{numbeqns}
\end{equation}
coupled quadratic equations for the same number of unknowns  (\ref{frange}).
 
The number of physically inequivalent solutions to the OVME expected$^{\rf{2}}$ at each level $\hat{k}$ is therefore
\begin{equation}
N(g,\lambda)\approx 2^{n(g,\lambda)-{\scriptsize \textup{dim}}g} \label{numbsolns}
\end{equation}
where we have subtracted the degrees of freedom associated with the Lie $g$ covariance$^{\rf{2}}$
\begin{equation}
\lr_r^{'ab}=\lr_r^{cd}(\omega^{-1})_c^{\ a}(\omega^{-1})_d^{\ b},\ \ \omega \in Aut(g)
\end{equation}
 of the OVME.  As examples of (\ref{numbsolns}), one finds
\alpheqn
\begin{eqnarray}
&&N(SU(2),\lambda=1)= 8,\hspace{.466in} N(SU(3),\lambda=1)\approx\frac{1}{4} \ \textrm{billion}\\
&&N(SU(2),\lambda=2)=512,\hspace{.3in} N(SU(3),\lambda=2)\approx 18 \ \textrm{quintillion}
\end{eqnarray}
\reseteqn
where $\lambda=1$ is the VME and the growth with $\lambda$ at fixed $g$ is exponential.

It is clear that, as in the VME, most of the solutions of the OVME will be new and (because the OVME is a large set of coupled quadratic equations) the new constructions will have generically irrational conformal weights and central charges.

\subsection{$\lr_{r}^{ab}$ independent of $r$}

For any $\lambda$, we consider the simple consistent ansatz
\begin{equation}
\lr_r^{ab}(\hat{G})=\lr^{ab}(\hat{G}),
\end{equation}
($\hat{G}$ is defined in (\ref{gdef})) which collects all inverse inertia tensors independent of $r$.  Then the OVME simplifies to a rescaled VME
\begin{equation}
\lr^{ab}(\hat{G}) =  \lambda[2 \lr^{ac}(\hat{G}) G_{cd} \lr^{db}(\hat{G}) - \lr^{cd}(\hat{G}) \lr^{ef}(\hat{G}) f_{ce}^{\ \ a}f_{df}^{\ \ b} - \lr^{cd}(\hat{G}) f_{ce}^{\ \ f}f_{df}^{\ \ (a}\lr^{b) e}(\hat{G})] \label{rindepOVME}
\end{equation}
where $G_{ab}$ is defined in (\ref{gabdef}).  Inspection of (\ref{rindepOVME}) gives
\namegroup{rindep}
\alpheqn
\begin{equation}
\hat{T}=\lr^{ab}(\hat{G})\sum_{r=0}^{\lambda-1}:\hat{J}_a^{(r)}\hat{J}_b^{(-r)}: \label{rindepT}
\end{equation}
\begin{equation}
\lr^{ab}(\hat{G})=\frac{1}{\lambda}L^{ab}(G) \label{rindeptensor}
\end{equation}
\begin{equation}
\hat{c}=\lambda c= 2 \lambda G_{ab}L^{ab}(G), \hspace{.2in} \hat{\Delta}_0=\frac{\hat{c}}{24}(1-\frac{1}{\lambda^2})
\end{equation}
\reseteqn
where $L^{ab}(G)$ is any solution of the VME (\ref{VMEa}) on semisimple affine Lie $g$. 

The constructions (\ref{rindep}) were given in Ref.~\rf{1}. They are the stress tensors of the twisted sectors of all the cyclic permutation orbifolds
\begin{equation}
\frac{A\stackrel{\lambda\hspace{.05in}times}{\times...\times}A}{\Z_{\lambda}}\label{corb}
\end{equation}
which can be constructed from $\lambda$ copies of any affine-Virasoro construction (\ref{VMEaa}).  In what follows, we will call (\ref{rindep}) the set of \textit{cyclic constructions}.\footnote{The stress tensors of the twisted sectors of the $S_n$ permutation orbifolds can also be constructed$^{\rf{25},\rf{1},\rf{23},\rf{24}}$ as sums of commuting copies of the cyclic constructions (\ref{rindep}).  This is because every element of $S_n$ can be expressed as a product of disjoint cyclic permutations.} 

Because they operate in cyclic permutation orbifolds, the cyclic constructions form a Virasoro subalgebra of the orbifold Virasoro algebra$^{\rf{1}}$
\alpheqn
\begin{equation}
\hat{T}^{(r)}(z)\hat{T}^{(s)}(w) = \frac{(\hat{c}/2)\delta_{r+s,0\hspace{.05in}mod\hspace{.03in}\lambda}}{(z-w)^4} + \frac{2 \hat{T}^{(r+s)}(w)}{(z-w)^2} + \frac{\partial_w\hat{T}^{(r+s)}(w)}{z-w}+O((z-w)^0)
\end{equation}
\begin{equation}\hat{T}^{(r)}(z)=\frac{1}{\lambda}L^{ab}(G)\sum_{s=0}^{\lambda-1}:\hat{J}_a^{(s)}(z)\hat{J}_b^{(r-s)}(z):\end{equation}
\reseteqn
where the zero-twist component $\hat{T}^{(0)}$ of the extended algebra is the cyclic construction $\hat{T}$ in (\ref{rindep}). 

Among the cyclic constructions, we note in particular the \textit{orbifold affine-Sugawara construction} $\hat{T}_{g_\lambda}$,
\namegroup{oassemi}
\alpheqn
\begin{equation}
\hat{T}_{g_\lambda}=\sum_{r=0}^{\lambda-1}(\lr_r^{ab})_{g_\lambda}:\hat{J}_a^{(r)}\hat{J}_b^{(-r)}:
\end{equation}
\begin{equation}
(\lr_{r}^{ab})_{g_\lambda}(\hat{G})=\frac{L^{ab}_g(G)}{\lambda}=\oplus_I\frac{\eta^{ab}_I}{2\hat{k}_I+\lambda Q_I}
\end{equation}
\begin{equation}
\hat{c}_{g_\lambda}=\lambda c_g= \lambda \sum_I \frac{2\hat{k}_I\textup{dim}g_I}{2\hat{k}_I+\lambda Q_I} ,\hspace{.2in}\hat{\Delta}^{g_\lambda}_0=\frac{\hat{c}_{g_\lambda}}{24}(1-\frac{1}{\lambda^2})
\end{equation}
\reseteqn
where $Q_I$ is the quadratic Casimir of $g_I$ and $c_g$ is the central charge of the affine-Sugawara construction$^{\rf{6},\rf{8}-\rf{14}}$ at levels $\{k_I\}$ on semisimple $g$.  The form of these constructions$^{\rf{4},\rf{5},\rf{1}}$ on simple $g$ is:
\namegroup{oassimple}
\alpheqn
\begin{equation}
(\lr_{r}^{ab})_{g_\lambda}(\hat{k})=\frac{L^{ab}_g(k)}{\lambda}=\frac{\eta^{ab}}{2\hat{k}+\lambda Q_g}
\end{equation}
\begin{equation}
\hat{c}_{g_\lambda}=\frac{\lambda\hat{x}\textup{dim}g}{\hat{x}+\lambda \tilde{h}_g}=\lambda c_g(x),\hspace{.2in}c_g(x)=\frac{x\textup{dim}g}{x+\tilde{h}_g}
\end{equation}
\begin{equation}
\hat{\Delta}^{g_\lambda}_0=\frac{\hat{c}_{g_\lambda}}{24}(1-\frac{1}{\lambda^2}). \label{oas}
\end{equation}
\reseteqn
Here $\tilde{h}_g=Q_g/\psi^2$ is the dual Coxeter number of $g$, $x\in\Z^+$ is the invariant level in (\ref{xdef}) and $c_g(x)$ is the central charge of the affine-Sugawara construction on simple $g$.

The orbifold affine-Sugawara constructions include the affine-Sugawara constructions $T_g$ at $\lambda=1$, and we shall see below that the orbifold affine-Sugawara constructions play the same fundamental role in the OVME that the affine-Sugawara constructions play in the VME.

\subsection{Constructions on subalgebras}

The OVME at order $\lambda$ has solutions $\lr(\lambda;h_\eta)$ given by
\namegroup{sub}
\alpheqn
\[\lr^{ab}_r(\lambda;h_\eta)=\left\{\begin{array}{l}\lr_s^{AB}(h_\eta), \hspace{.5in}\textrm{if}\ \exists s\in [0,\eta -1]\ \textrm{s.t.}\ r=\frac{\lambda}{\eta}s\\
0,\hspace{1in}\textrm{otherwise}\ 
\end{array}\right.\]
\vspace{-.4in}
\begin{equation}\end{equation}
\vspace{-.2in}
\begin{equation}A,B=1,...,\textup{dim}h,\hspace{.2in}r=0,...,\lambda-1,\hspace{.2in}\eta\in\Z^+,\hspace{.2in}\frac{\lambda}{\eta}\in{\Z^+}
\end{equation}
\begin{equation}
\hat{c}(\lambda;h_\eta) = \hat{c}(h_\eta),\hspace{.2in} \ \hat{\Delta}_0(\lambda;h_\eta)=\hat{\Delta}_0(h_\eta)
\end{equation}
\reseteqn
where $\lr(h_\eta)$ is any solution to the OVME on $h_\eta$, $h\subset g$ (with central charge $\hat{c}(h_\eta)$ and ground state conformal weight $\hat{\Delta}_0(h_\eta)$).  The order $\eta$ solutions appear at order $\lambda$ because $h_\eta\subset g_\lambda$ (see Section~\ref{sec-sub}).

The subalgebra constructions (\ref{sub}) include in particular the orbifold affine-Sugawara construction on $h_\eta\subset g_\lambda$,
\namegroup{th}
\alpheqn
\begin{equation}
\hat{T}_{h_\eta}(\lambda;h_\eta)=\frac{\eta^{AB}}{2\hat{k}+\eta Q_h}\sum_{r=0}^{\eta-1}:\hat{J}_A^{(\frac{\lambda}{\eta}r)}\hat{J}_B^{(-\frac{\lambda}{\eta}r)}:,\hspace{.2in}
A,B=1,...,\textup{dim}h
\end{equation}
\begin{equation}
\hat{c}_{h_\eta}=\frac{r\eta\hat{x}\textup{dim}h}{r\hat{x}+\eta \tilde{h}_h}
,\hspace{.2in}
\hat{\Delta}^{h_\eta}_0=\frac{\hat{c}_{h_\eta}}{24}(1-\frac{1}{\eta^2})
\end{equation}
\reseteqn
for simple $g$ and simple $h$ where $r$ is the index of embedding of $h\subset g$.

We also mention the case $h_\eta=g_1$ of (\ref{sub}), which collects all constructions on the integral affine subalgebra:
\alpheqn
\begin{equation}
\lr_r^{ab}(\lambda;g_1) = \delta_{r,0} L^{ab}(\hat{G})
\end{equation}
\begin{equation}
\hat{T}(\lambda;g_1) = L^{ab}(\hat{G}):\hat{J}_a^{(0)}\hat{J}_b^{(0)}:\hspace{.05in} , \hspace{.3in} \hat{c}(\lambda;g_1) = 2L^{ab}(\hat{G})\hat{G}_{ab}, \hspace{.3in} \hat{\Delta}_0(\lambda;g_1)=0
\end{equation}
\reseteqn
where $L^{ab}(\hat{G})$ is any solution to the VME (\ref{VMEa}) with $G_{ab}$ replaced by $\hat{G}_{ab}$ in (\ref{gdef}).  These stress tensors are isomorphic to the stress tensors of the VME and the ground state conformal weights vanish, as expected, for all these constructions because $\lr_0^{ab}$ does not contribute to $\hat{\Delta}_0$ in (\ref{delta}).

\subsection{K-conjugation covariance}
\label{ksec}

At order $\lambda$, the OVME exhibits \textit{K-conjugation covariance} through the orbifold affine-Sugawara construction $\hat{T}_{g_\lambda}$, which includes the familiar K-conjugation covariance$^{\rf{6},\rf{8},\rf{7},\rf{15},\rf{9}}$ through the affine-Sugawara construction when $\lambda=1$.  This means (see App. A) that the K-conjugate inertia tensor $\tilde{\lr}$
\begin{equation}
\tilde{\lr}_r^{ab}=(\lr_r^{ab})_{g_\lambda}-\lr_r^{ab}
\end{equation}
is a solution of the OVME when $\lr$ is a solution, and 
\alpheqn
\begin{equation}
\tilde{\hat{T}\hspace{.035in}}=\hat{T}_{g_\lambda}-\hat{T},\hspace{.2in}\tilde{\hat{c}\hspace{.018in}}=\hat{c}_{g_{\lambda}}-\hat{c},\hspace{.2in}\tilde{\hat{\Delta\hspace{.01in}}}_0=\hat{\Delta}^{g_\lambda}_0-\hat{\Delta}_0, \end{equation}
\begin{equation}
\hat{T}(z)\tilde{\hat{T}\hspace{.025in}}(w)=O((z-w)^0) \label{commute}
\end{equation}
\reseteqn
where $\tilde{\hat{T}\hspace{.035in}}$ is the corresponding K-conjugate stress tensor.

Drawing on experience with the VME, K-conjugation covariance tells us that the conformal field theory corresponding to $\hat{T}$ can be considered as a gauge theory$^{\rf{29},\rf{30},\rf{2},\rf{19}-\rf{22}}$ in which the cyclic permutation orbifold of $\lambda$ copies of WZW is gauged by the K-conjugate partner $\tilde{\hat{T}\hspace{.035in}}\hspace{.02in}$of $\hspace{.05in}\hat{T}$.

The OVME has another covariance, which we call $Aut({\Z}_\lambda)$ covariance, that has no analog in the VME:  For any automorphism of ${\Z}_\lambda$, the inertia tensor $\lr^{'}$
\begin{equation}\lr_r^{'ab}=\lr_{\phi(r)}^{ab},\hspace{.3in} \hat{c}^{'} =\hat{c}, \hspace{.3in}\phi\in Aut({\Z}_\lambda)
\end{equation}
is a solution when $\lr$ is a solution.  $Aut({\Z}_\lambda)$ covariance relates conformal constructions with the same central charge but generically different conformal weights. 

\subsection{Coset constructions}

At order $\lambda$, consider the orbifold affine-Sugawara construction $\hat{T}_{h_\eta}\equiv\hat{T}_{h_\eta}(\lambda;h_\eta)$ on $h_\eta\subset g_\lambda$ where $g$ and $h$ are semisimple.  (For simple $g$ and $h$, this construction is given in (\ref{th})).  The K-conjugate partner of $\hat{T}_{h_\eta}$ is the general $g_{\lambda}/h_{\eta}$ coset construction,
\begin{equation}
\hat{T}_{g_\lambda/h_\eta}=\tilde{\hat{T}\hspace{.035in}} = \hat{T}_{g_\lambda}-\hat{T}_{h_\eta},
\hspace{.3in}\hat{c}_{g_\lambda/h_\eta}=\tilde{\hat{c}\hspace{.018in}}=\hat{c}_{g_\lambda}-\hat{c}_{h_\eta} \label{coset}
\end{equation}
which includes the ordinary coset constructions$^{\rf{6},\rf{8},\rf{7}}$ at $\lambda=1$.  The special case of (\ref{coset}) given by Ka\u{c} and Wakimoto$^{\rf{4},\rf{5}}$ is $g_\lambda/g_{\eta=1}$, that is $\eta=1$ and $h=g$ with
\begin{equation}
\hat{c}_{g_\lambda/g_{\eta=1}}=\hat{c}_{g_\lambda/g_1}=\lambda c_g(x)-c_g(\lambda x).
\end{equation}
The extension to $g_{\lambda}/h_{\eta=1}$ was given in Ref.~\rf{1}.

It was conjectured in Ref.~\rf{1} that the coset construction $g_{\lambda}/h_{\eta=1}$ corresponds to the twisted sectors of a different kind of orbifold
\begin{equation}
\frac{(\frac{g\stackrel{\lambda\hspace{.05in}times}{\times...\times}g}
{h_D(\lambda)})}{{\Z}_{\lambda}}
\end{equation}
where $h_D(\lambda)$ is the diagonal subalgebra of $h \times...\times h$.

\subsection{Nests}

Repeated K-conjugation on nested orbifold affine subalgebras $g_\lambda\supset h^{(1)}_{\eta_1}\supset...\supset h^{(n)}_{\eta_n}$ gives the \textit{orbifold affine-Sugawara nests}, which generalize the affine-Sugawara nests$^{\rf{34},\rf{16},\rf{2}}$ of the VME on $g\supset h^{(1)}\supset...\supset h^{(n)}$.  The orbifold affine-Sugawara constructions and the coset constructions are the lowest nests, and the first non-trivial nests are 
\alpheqn
\begin{equation}
\hat{T}_{g_\lambda/h_\eta/h^\prime_{\eta^\prime}} = \hat{T}_{g_\lambda} - (\hat{T}_{h_\eta}-\hat{T}_{h^\prime_{\eta^\prime}})
\end{equation}
\begin{equation}
\hat{c}_{g_\lambda/h_\eta/h^\prime_{\eta^\prime}} = \hat{c}_{g_\lambda} - (\hat{c}_{h_\eta}-\hat{c}_{h^\prime_{\eta^\prime}}).
\end{equation}
\reseteqn
More generally one has the \textit{orbifold affine-Virasoro nests}, including
\alpheqn
\begin{equation}\hat{T}^\#_{g_\lambda/h_\eta/h^\prime_{\eta^\prime}} = \hat{T}_{g_\lambda} - (\hat{T}_{h_\eta}-\hat{T}^\#_{h^\prime_{\eta^\prime}})
\end{equation}
\begin{equation}\hat{c}^\#_{g_\lambda/h_\eta/h^\prime_{\eta^\prime}} = \hat{c}_{g_\lambda} - (\hat{c}_{h_\eta}-\hat{c}^\#_{h^\prime_{\eta^\prime}})\end{equation}
\reseteqn
where $\hat{T}^\#_{h^\prime_{\eta^\prime}}$ is an arbitrary construction on the subalgebra $h^\prime_{\eta^\prime}$.  These nests generalize the affine-Virasoro nests$^{\rf{16},\rf{2}}$ familiar at $\lambda=1$.

In what follows, we refer to the solutions discussed above (the cyclic constructions, the subalgebra constructions on $h_\eta\subset g_\lambda$ and the orbifold affine-Virasoro nests) as ``known'' solutions of the OVME and all other solutions will be called ``new''.

\section{The Lie $g$-Invariant Constructions}

Many consistent ans$\ddot{\textrm{a}}$tze$^{\rf{16},\rf{2}}$ and subans$\ddot{\textrm{a}}$tze can be found for the OVME, as for the VME.  In this section we concentrate on the Lie $g$-invariant constructions on simple $g$, whose abundance is a surprising feature of the OVME.  The generalization of the ``graph theory ansatz'' on $SO(n)$, familiar$^{\rf{28}}$ at $\lambda=1$, is also given in App.~B.
\subsection{Group-invariant ansatz}
\setcounter{equation}{0}
We consider the group-invariant ansatz
\begin{equation}\psi^2\lr_r^{ab}=l_r \eta^{ab},\ \ \ r=0,...,\hl \label{ginvar}
\end{equation}
($\eta^{ab}$ is the inverse Killing metric of $g$) which collects all \textit {Lie $g$-invariant constructions} in the OVME on simple $g$.  We know that this ansatz includes at least the trivial construction $\lr=0$, the orbifold affine-Sugawara construction $\hat{T}_{g_\lambda}$, the particular subalgebra constructions $\hat{T}_{g_{\eta}}$, the coset constructions $g_\lambda /g_\eta$ and all orbifold affine-Sugawara nests of the form $g_\lambda\supset g_\eta \supset...\supset g_{\eta\prime}$.  Among these only two constructions survive at $\lambda=1$, namely the trivial construction and the affine-Sugawara construction on $g$.  We shall see that the number of new Lie $g$-invariant constructions increases rapidly with $\lambda$.
 
Substitution of (\ref{ginvar}) into (\ref{OVME}) gives the reduced OVME of the ansatz:
\namegroup{OVMELie}
\alpheqn
\begin{equation}
\psi^2 \hat{T}=\sum_{r=0}^{\lambda-1}l_r\eta^{ab}:\hat{J}_a^{(r)}\hat{J}_b^{(-r)}:
\end{equation}
\begin{equation}
l_r  = \hat{x} l_r^{\ 2} + \tilde{h}_g\sum^{\lambda-1}_{s=0} l_s (2 l_r -  l_{r+s}),\hspace{.2in}0\leq r\leq\hl \label{g}
\end{equation}
\begin{equation}
l_r=l_{r\pm\lambda}=l_{\lambda\pm r},\hspace{.2in}\lfloor\hle\rfloor<r\leq\lambda
\end{equation}
\begin{equation}
\hat{c}=\hat{x}\textup{dim}g(\sum_{r=0}^{\lambda-1}l_r),\hspace{.2in}\hat{\Delta}_0=\hat{x}\textup{dim}g\sum_{r=0}^{\lambda-1}l_r\frac{r(\lambda-r)}{4\lambda^2}
\end{equation}
\reseteqn
where $\psi$ is the highest root of $g$ and $\hat{x}=\lambda x,\ x\in\Z^+$.  For computational purposes we give also the forms of the reduced OVME in the fundamental range:

\underline{For even $\lambda$}
\namegroup{ge}
\alpheqn
\begin{equation}
\psi^2\hat{T} = 
 l_0\eta^{ab}:\hat{J}_a^{(0)}\hat{J}_b^{(0)}:+l_{\lambda/2}\eta^{ab}:\hat{J}_a^{(\lambda/2)}\hat{J}_b^{(-\lambda/2)}:+2\sum_{r=1}^
{\hle-1}l_r\eta^{ab}:\hat{J}_a^{(r)}\hat{J}_b^{(-r)}:
\end{equation}
\vspace{-.2in}
\begin{eqnarray}
l_r  & = &  \hat{x} l_r^{\ 2} - 
\tilde{h}_g(2\sum^{\hle-r-1}_{s=0} l_sl_{r+s}
+\sum^{\hle}_{s=\hle-r} l_s l_{\lambda-r-s}
+\sum^{r}_{s=0} l_s l_{r-s}-2l_0l_r)\nonumber\\ & &\ \  +2\tilde{h}_gl_r(l_0 +  l_{\lambda/2} + 2 \sum_{s=1}^{\frac{\lambda}{2}-1}l_s ),\hspace{.4in}r=0,...,\hle \label{geb}
\end{eqnarray}
\vspace{-.2in}
\begin{equation}
\hat{c}=\hat{x}\textup{dim}g(l_0+l_{\lambda/2}+2\sum_{r=1}^{\frac{\lambda}{2}-1}l_r),\hspace{.2in}\hat{\Delta}_0=\hat{x}\textup{dim}g(\frac{l_{\lambda/2}}{16}+\sum_{r=1}^{\hle-1}l_r\frac{r(\lambda-r)}{2\lambda^2}).
\end{equation}
\reseteqn
\underline{For odd $\lambda$}
\namegroup{go}
\alpheqn
\begin{equation}
\psi^2\hat{T} = 
 l_0\eta^{ab}:\hat{J}_a^{(0)}\hat{J}_b^{(0)}:+2\sum_{r=1}^{\hlo}l_r\eta^{ab}:\hat{J}_a^{(r)}\hat{J}_b^{(-r)}:
\end{equation}
\vspace{-.2in}
\begin{eqnarray}
l_r  & = &  \hat{x} l_r^{\ 2} - 
\tilde{h}_g(2\sum^{\hlo-r}_{s=1} l_sl_{r+s}
+\sum^{\hlo}_{s=\hlo+1-r} l_s l_{\lambda-r-s}
+\sum^{r}_{s=0} l_s l_{r-s})\nonumber\\ & &\ \  +2\tilde{h}_gl_r(l_0 + 2 \sum_{s=1}^{\hlo}l_s ),\hspace{.4in}r=0,...,\hlo \label{gob}
\end{eqnarray}
\vspace{-.2in}
\begin{equation} 
\hat{c} =\hat{x}\textup{dim}g(l_0+2\sum_{r=1}^{\hlo}l_r),\hspace{.2in}\hat{\Delta}_0=\hat{x}\textup{dim}g(\sum_{r=1}^{\hlo}l_r\frac{r(\lambda-r)}{2\lambda^2}).
\end{equation}
\reseteqn
We see that the group-invariant ansatz (\ref{ginvar}) is a consistent ansatz, giving $\lfloor \frac{\lambda}{2} \rfloor +1$ coupled quadratic equations and unknowns.  
 
At each level $\hat{x}$, it follows that there are 
\begin{equation}
N(\lambda) = 2^{\lfloor \frac{\lambda}{2} \rfloor + 1} 
\end{equation}
Lie $g$-invariant solutions at order $\lambda$.  In fact, these solutions are organized into $N(\lambda)$ level families$^{\rf{2}}$, according to the high-level behavior$^{\rf{28},\rf{2}}$
\begin{equation}
l_r=\frac{\theta_r}{\hat{x}}+O(\hat{x}^{-2}),\hspace{.3in}\theta_r\in\{0,1\}
,\hspace{.3in}r=0,...,\hl
\end{equation}
of each level family.  Only four of these level families are ``known'' at prime $\lambda$, and so there will be
\begin{equation}
N_{new}(\lambda)=2^{\hl+1}-4, \hspace{.3in} \lambda \ {\textrm{prime}}
\end{equation}
new Lie $g$-invariant level families at these orders.
 
The high-level expansion$^{\rf{28},\rf{2}}$ of the Lie $g$-invariant level families also shows that each level family is unitary for $\hat{x}=\lambda x$, $x\in\Z^+$, at least down to some finite radius of convergence $\hat{x}_0=\lambda x_0$.  Experience$^{\rf{2}}$ with the VME shows that $x_0$ is usually quite low, for example $x_0=1$ or $2$ (see also Subsecs. \ref{lambdafive} and \ref{lambdasix}).

We turn now to finding exact solutions of Eqs.~(\ref{geb}) and (\ref{gob}).  In solving these equations it was useful to note that only the $r=0$ equations have $l_0$ dependence on their right hand sides, which effectively reduces the number of equations by one.  Then we were able to factorize$^{\rf{16},\rf{2}}$ linear combinations of the $r\neq0$ equations through $\lambda=6$.  We find no new solutions \footnote{For example  the four solutions at $\lambda=2$ are the trivial construction, the orbifold affine-Sugawara construction, the affine-Sugawara construction on the integral affine subalgebra and a Ka\u{c}-Wakimoto coset.} for $\lambda=1$, 2, 3 and 4.  The results for $\lambda=5$ and $6$ are reported below.  

\subsection{Irrational conformal weights at $\lambda = 5$}
\label{lambdafive}

At $\lambda=5$ there are $2^{2+1}=8$ level families, four of which are new:
\alpheqn
\begin{eqnarray}
l_0 & = & \frac{1}{2(\hat{x} + 5 \tilde{h}_g)} + \frac{\theta}{2(\hat{x} + \tilde{h}_g)}\\
l_1 & = & \frac{1}{2(\hat{x} + 5 \tilde{h}_g)}(1 + \eta\sqrt{\frac{\hat{x} + 5\tilde{h}_g}{\hat{x} + \tilde{h}_g}})\\
l_2 & = & \frac{1}{2(\hat{x} + 5 \tilde{h}_g)}(1 - \eta\sqrt{\frac{\hat{x} + 5\tilde{h}_g}{\hat{x} + \tilde{h}_g}})\\
\hat{c} &=& \frac{\hat{x}}{2} \ \textup{dim}g(\frac{5}{\hat{x} + 5 \tilde{h}_g} + \frac{\theta }{\hat{x} + \tilde{h}_g}), \hspace{.3in} \eta, \theta= \  \pm 1\ . \label{eq:c} 
\end{eqnarray}
\reseteqn
K-conjugation takes $(\theta,\eta) \rightarrow (-\theta,-\eta)$ and all these constructions are unitary \footnote{It also follows that the high-level expansions of these level families are convergent down to and including level $x=1$.} for $\hat{x} =5x$, $x \in \Z^+$ because the $\lr$'s are real$^{\rf{2}}$.
 
The central charges (\ref{eq:c}) are rational, but the conformal weights of the ground states of these constructions are generically irrational  
\begin{equation}
\hat{\Delta}_0 = \frac{\hat{x} \textup{dim}g}{10} (\frac{1}{\hat{x} + 5 \tilde{h}_g} - \frac{\eta}{5 \sqrt{(\hat{x} + 5 \tilde{h}_g)(\hat{x} + \tilde{h}_g)}}) \label{cw5}
\end{equation}
and other conformal weights will be generically irrational because the $\lr$'s are irrational.  We will discuss this set of new constructions further in Sec. 6.

\subsection{Unitary irrational central charge at $\lambda=6$}
\label{lambdasix}

At $\lambda =6$ there are $2^{3+1}=16$ level families, four of which are new: 
\alpheqn
\begin{eqnarray}
l_0 & = & \frac{1}{2(\hat{x} + 6 \tilde{h}_g)} ( 1 + \eta \frac{\tilde{h}_g(- \hat{x}^2 + 3 \hat{x} \tilde{h}_g - 6 \tilde{h}_g^2)}{(\hat{x} + \tilde{h}_g) \alpha}) + \frac{\theta}{2(\hat{x} + \tilde{h}_g)} \\
l_1  & = & \frac{1}{2(\hat{x} + 6 \tilde{h}_g)} ( 1 + \eta \frac{- \hat{x}^2 - 5 \hat{x} \tilde{h}_g + 18 \tilde{h}_g^2}{\alpha}) \\
l_2 & = & \frac{1}{2(\hat{x} + 6 \tilde{h}_g)} ( 1 + \eta \frac{\hat{x}^2 + 3 \hat{x} \tilde{h}_g - 6 \tilde{h}_g^2}{\alpha}) \\
l_3 & = & \frac{1}{2(\hat{x} + 6 \tilde{h}_g)} ( 1 + \eta \frac{\hat{x}^2 + \hat{x} \tilde{h}_g - 18 \tilde{h}_g^2}{\alpha})\\ 
\alpha &\equiv& \sqrt{\hat{x}^4 + 2 \hat{x}^3 \tilde{h}_g -19 \hat{x}^2 \tilde{h}_g^2 + 12 \hat{x} \tilde{h}_g^3 + 36 \tilde{h}_g^4}, \hspace{.3in} \eta, \theta= \pm 1. \label{alpha}
\end{eqnarray}
\reseteqn
As in the previous case, K-conjugation takes $(\theta,\eta)\rightarrow (-\theta,-\eta)$.
 
Using the fact that the dual Coxeter number $\tilde{h}_g$ is positive for all compact Lie algebras, we have checked that $\alpha$, and hence $\lr_r^{ab}$, are real for all levels $\hat{x}=6x$, $x\in\Z^+$ of all compact Lie algebras.  It follows that these constructions are also unitary\footnote{The high-level expansions of these level families are also convergent through level $x=1$.} for these levels and algebras.
 
For these constructions, both the central charge and the ground state conformal weight are generically irrational:
\alpheqn
\begin{eqnarray}
\hat{c}&=&\frac{\hat{x}}{2} \ \textup{dim}g(\frac{6}{\hat{x}+ 6\tilde{h}_g} + \eta \frac{\hat{x}^3 - 3 \hat{x}^2 \tilde{h}_g + 6 \hat{x}\tilde{h}_g^2}{\alpha (\hat{x}+6\tilde{h}_g)(\hat{x}+\tilde{h}_g)}+ \frac{\theta}{\hat{x} +\tilde{h}_g}) \label{c6}\\
\hat{\Delta}_0&=& \frac{\hat{x} \ \textup{dim}g}{288(\hat{x} + 6 \tilde{h}_g)} (35 + \eta \frac{15 \hat{x}^2 + 7 \hat{x} \tilde{h}_g - 78 \tilde{h}_g^2}{\alpha})
\end{eqnarray}
\reseteqn
due to the quantity $\alpha$ in (\ref{alpha}).  For certain levels the central charge is rational (for example level $\hat{x}=6x=6$ of $SU(2)$ and $SU(3)$).
 
The lowest irrational central charges for each of the simple Lie algebras occur at $\theta=\eta=-1$ and are listed below:

\vspace{.3in}

\hspace{1.766in}\begin{tabular}{r|l|l}
&$\hat{x}$&$\hspace{.9in}\hat{c}$\\ \hline
$SU(2)$ & $12 $&$ \hspace{.2in} \frac{9}{14}(5-\frac{1}{\sqrt{2}}) \approx 2.7597$ \\
$SO(3)$ & $12 $&$ \hspace{.2in}\frac{9}{14}(5-\frac{1}{\sqrt{2}}) \approx 2.7597$ \\
$SP(1)$ & $12 $&$ \hspace{.2in}\frac{9}{14}(5-\frac{1}{\sqrt{2}}) \approx 2.7597$ \\
$E_6$  & $6 $&$ \hspace{.2in}5- \frac{19}{\sqrt{601}} \approx 4.2249$ \\
$E_7$  & $6 $&$ \hspace{.2in}\frac{35}{8} - \frac{161}{8 \sqrt{769}} \approx 3.6492$ \\
$E_8$  & $6 $&$ \hspace{.2in}\frac{10}{3}-\frac{68}{3\sqrt{1471}} \approx 2.7423$ \\
$F_4$  & $6 $&$ \hspace{.2in}\frac{26}{5} - \frac{26}{5 \sqrt{46}} \approx 4.4330$ \\
$G_2$  & $6 $&$ \hspace{.2in}\frac{21}{5} - \frac{21}{5\sqrt{41}} \approx 3.5440$ \\
\end{tabular}
\[\]
At fixed $g$, the central charges increase monotonically with $\hat{x}$, and, for the classical Lie algebras (e.g. $SU(n)$), the central charges increase monotonically with $n$.

The lowest unitary irrational central charge in this family of constructions is therefore
\begin{equation}
\hat{c}( (E_8)_{\hat{x}=6},\lambda=6)=\frac{10}{3}-\frac{68}{3\sqrt{1471}} \approx 2.7423 \label{eq:c2}
\end{equation}  
and we note that level $\hat{x}=6$ of $E_8$ corresponds, via the orbifold induction procedure$^{\rf{1}}$, to the historic level $x=1$ of $E_8$.

The value (\ref{eq:c2}) is less than the lowest known unitary irrational central charge among the cyclic permutation orbifolds (\ref{corb}) at $\lambda=6$,
\begin{equation}
\hat{c}=6 \ c(SU(3)_5)^{\#}_{D(1)})=6 \ (2)(1-\frac{1}{\sqrt{61}}) \approx 10.4635
\end{equation}
where $c(SU(3)_5)^{\#}_{D(1)})$ is the lowest known$^{\rf{2}}$ unitary irrational central charge in the VME.  The central charge in (\ref{eq:c2}) is also less than $\hat{c}=6\times 1=6$, which is the lower bound on possible unitary irrational central charge of any order $6$ cyclic permutation orbifold.

\section{Discussion: The Lie $h$-invariant constructions}
\setcounter{equation}{0}

The solutions above for $\lambda \leq 6$ show that the Lie $g$-invariant constructions of the OVME are closely related to the Lie $h$-invariant constructions$^{\rf{17},\rf{2}}$, with $h \subset g$, which have been studied in the VME.  Following Ref.~\rf{17}, the Lie $h$-invariant constructions of the OVME are those constructions on $g_{\lambda}$ whose inertia tensors are Lie $h$-invariant 
\begin{equation}
\delta \lr_r^{ab}=\lr_r^{c(a}f_{cd}^{\ \ b)} \psi^d=0, \hspace{.3in} r=0,...,\lfloor \hle \rfloor
\end{equation}
where $\psi^a$ parameterizes the Lie group $H \subset G$ near the origin.  Then we know from Ref.~\rf{17} that these constructions occur in Lie $h$-invariant quartets, octets, etc., and the currents of the Lie $h$-invariant stress tensors are either $(1,0)$ operators (associated to a global $h$-symmetry) or $(0,0)$ operators (associated to a local $h$-symmetry).
 
In the case of the OVME, Lie $h$ symmetry guarantees at least a global or local $h_1$ symmetry, although $h_{\eta}$ symmetry can also appear.  For example, the general coset construction on $g_{\lambda}/h_{\eta}$ lives in the Lie $h$-invariant quartet

\begin{picture}(350,185)(0,0)
\put(135,150){$\hat{T}_{g_\lambda /h_\eta}$}
\put(170,155){\vector(1,0){95}}
\put(205,160){$+\hat{T}_{h_\eta}$}
\put(275,150){$\hat{T}_{g_\lambda}$}
\put(150,140){\vector(0,-1){80}}
\put(150,60){\vector(0,1){80}}
\put(125,105){$K_{g_\lambda}$}
\put(165,140){\vector(4,-3){110}}
\put(271,60){\vector(-4,3){110}}
\put(223,105){$K_{g_\lambda /h_\eta}$}
\put(283,140){\vector(0,-1){80}}
\put(283,60){\vector(0,1){80}}
\put(290,105){$K_{g_\lambda}$}
\put(145,45){$\hat{T}_{h_\eta}$}
\put(270,50){\vector(-1,0){100}}
\put(205,55){$+\hat{T}_{h_\eta}$}
\put(280,45){$0$}
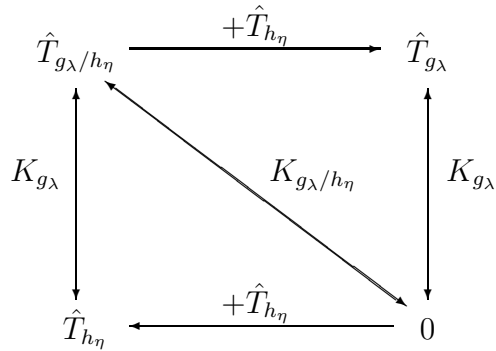
\put(100,15){Figure 1: The simplest Lie $h$-invariant quartet}
\end{picture}\\
where $K_{g_{\lambda}}$ is the usual K-conjugation $(\tilde{\hat{T}\hspace{.035in}} = \hat{T}_{g_\lambda}-\hat{T})$ through the orbifold affine-Sugawara construction on $g_{\lambda}$ and $K_{g_{\lambda}/h_{\eta}}$ is (a trivial example of) another K-conjugation through the coset construction itself ($\tilde{\hat{T}\hspace{.035in}} = \hat{T}_{g_\lambda/ h_\eta}-\hat{T}$ with $\hat{T}=0$ or $\hat{T}_{g_\lambda/ h_\eta})$.

The general Lie $h$-invariant quartet has the form

\begin{picture}(350,185)(0,0)
\put(120,150){$\hat{T}(h_\eta \subset g_\lambda)$}
\put(190,155){\vector(1,0){50}}
\put(200,160){$+\hat{T}_{h_\eta}$}
\put(250,150){$\hat{T}(h_\eta \subset g_\lambda)+\hat{T}_{h_\eta}$}
\put(150,140){\vector(0,-1){80}}
\put(150,60){\vector(0,1){80}}
\put(125,105){$K_{g_\lambda}$}
\put(165,140){\vector(4,-3){110}}
\put(271,60){\vector(-4,3){110}}
\put(223,105){$K_{g_\lambda /h_\eta}$}
\put(283,140){\vector(0,-1){80}}
\put(283,60){\vector(0,1){80}}
\put(290,105){$K_{g_\lambda}$}
\put(100,45){$\hat{T}_{g_\lambda}-\hat{T}(h_\eta \subset g_\lambda)$}
\put(244,50){\vector(-1,0){53}}
\put(204,55){$+\hat{T}_{h_\eta}$}
\put(250,45){$\hat{T}_{g_\lambda/ h_\eta}-\hat{T}(h_\eta \subset g_\lambda)$}
\put(100,15){Figure 2: The general Lie $h$-invariant quartet}
\end{picture}\\
where $\hat{T}(h_{\eta} \subset g_{\lambda})$ is any locally Lie $h$-invariant construction on $g_{\lambda}$ (so that $\hat{T}$ commutes with the currents of $h_{\eta}$).  The K-conjugation $K_{g_\lambda /h_\eta}$ through the coset construction relates the two locally-invariant constructions in the quartet, the other two constructions being globally invariant.  For the Lie $g$-invariant ansatz, one should read $h_{\eta} \rightarrow g_{\eta}$.

We have checked that the two new sets of constructions at $\lambda=5$ and $6$ are Lie $g$-invariant quartets with the form 

\begin{picture}(350,185)(0,0)
\put(103,150){$\eta=+1,\ \theta=-1$}
\put(191,155){\vector(1,0){50}}
\put(200,160){$+\hat{T}_{g_1}$}
\put(245,150){$\eta=+1,\ \theta=+1$}
\put(150,140){\vector(0,-1){80}}
\put(150,60){\vector(0,1){80}}
\put(125,105){$K_{g_\lambda}$}
\put(165,140){\vector(4,-3){110}}
\put(271,60){\vector(-4,3){110}}
\put(223,105){$K_{g_\lambda /g_1}$}
\put(283,140){\vector(0,-1){80}}
\put(283,60){\vector(0,1){80}}
\put(290,105){$K_{g_\lambda}$}
\put(103,45){$\eta=-1,\ \theta=+1$}
\put(244,48){\vector(-1,0){53}}
\put(204,55){$+\hat{T}_{g_1}$}
\put(250,45){$\eta=-1,\ \theta=-1$}
\put(100,15){Figure 3: New Lie $g$-invariant quartets at $\lambda=5,6$}
\end{picture}\\
so that $K_{g_\lambda /g_1}$ is  conjugation through the Ka\u{c}-Wakimoto coset construction in both these cases.  The constructions with $\theta=-1$ have a local $g$-invariance while the constructions with $\theta=1$ have only a global $g$-invariance.

To check these conclusions for all the Lie $g$-invariant constructions, we have evaluated the $\hat{T}\hat{J}^{(0)}$ OPE in this case to find 
\alpheqn
\begin{equation}
\hat{T}(z)\hat{J}_a^{(0)}(w)=M(\lr)_a^{\ b} (\frac{1}{(z-w)^2}+\frac{\partial_w}{z-w})\hat{J}_b^{(0)}+O((z-w)^0)
\end{equation}
\begin{equation}
M(\lr)_a^{\ b}=\Delta\delta_a^{\ b}, \hspace{.3in} \Delta \equiv \hat{x}l_0+\tilde{h}_g\sum_{r=0}^{\lambda-1}l_r \label{DEL}
\end{equation}
\reseteqn
where the matrix $M(\lr)_a^{\ b}$ is defined for all $\lr$ in App. A.
We have also checked that
\begin{equation}
\Delta^2=\Delta \label{nil}
\end{equation}
using the reduced OVME (\ref{g}) of the Lie $g$-invariant ansatz.  The result (\ref{nil}) verifies that the currents $\hat{J}_a^{(0)}$ of the integral affine subalgebra $g_{\eta=1}$ are either $(1,0)$ or $(0,0)$ operators in all Lie $g$-invariant constructions, as they should be.  The specific identifications in Fig.~$3$ follow easily from the central charges in (\ref{eq:c}) and (\ref{c6}).

Although all the Lie $g$-invariant constructions have a local or global symmetry associated to $g_{\eta=1}$, it is also possible to have larger symmetries associated to $h_\eta=g_{\eta}$, $\eta\geq 2$.  Indeed, there is a quartet at $\lambda=4$ associated to $g_{\eta=2}$.  One remaining question is the nature of the extra (perhaps discrete) symmetry of $\lr_r^{ab}$ which dictates the larger symmetry.

In the case of the $\lambda=5$ quartet, one sees further structure because there are only two independent central charges in (\ref{eq:c}).  In fact, we may write these central charges as
\[
\hat{c}=\frac{1}{2}(\hat{c}_{g_{\lambda=5}}+\theta \hat{c}_{g_{\eta=1}})=\left\{\begin{array}{cl}
\frac{1}{2}(\hat{c}_{g_5}+\hat{c}_{g_1})\\
\frac{1}{2}\hat{c}_{g_5/g_1}\\
\end{array}\right.
\]
\vspace{-.4in}
\begin{equation} \label{cquart} \end{equation}
which shows that the two $g_1$-local theories $(\theta=-1)$ have the same central charge, and moreover that this central charge is exactly half of the central charge of the Ka\u{c}-Wakimoto coset.

Local Lie $h$-invariant constructions with $c=\frac{1}{2}c_{g/h}$, called the self $K_{g/h}$-conjugate constructions \hspace{-.05in}$^{\rf{17},\rf{2}}$, are known from the VME, where $\tilde{\hat{T}\hspace{.035in}}$ and $\hat{T}$ in $\tilde{\hat{T}\hspace{.035in}} = \hat{T}_{g_\lambda}-\hat{T}$ are automorphically equivalent inertia tensors (related by a transformation in $Aut(h)$).  The mechanism here, although similar, is not the same:  In the present case, conjugation by $K_{g_5/g_1}$ $(\eta\rightarrow -\eta)$ is an example of the $Aut(\Z_\lambda)$ covariance of the OVME (see Subsec.~\ref{ksec}), so that the two constructions with $\hat{c}=\frac{1}{2}\hat{c}_{g_5/g_1}$ have generically different conformal weights.

The central charges in (\ref{eq:c}) (or (\ref{cquart})) can also be written as
\begin{equation}
\hat{c}=5(\frac{c_g(x)}{2})+\theta (\frac{c_g(5x)}{2}) \label{orbc5}, \hspace{.2in} c_g(x)=\frac{x \textup{dim}g}{x+\tilde{h}_g}, \hspace{.3in} x\in\Z^+
\end{equation}
where $c_g(x)$ is the central charge of the affine-Sugawara construction on $g$.  Since conventional orbifoldization does not change the central charge of a theory, the form (\ref{orbc5}) suggests that the new constructions at $\lambda=5$ might be twisted sectors of orbifolds which start with copies of a conformal field theory whose central charge is $c=c_g/2$.  Conformal field theories with $c=c_g/2$ are in fact known (the self K-conjugate constructions$^{\rf{28},\rf{2}}$ at $\lambda=1$), but these occur only for $\textup{dim}g= \textup{even}$, leaving us without a conventional orbifold interpretation for the new constructions at $\lambda=5$.  For the new constructions at $\lambda=6$, we have no suggestion at present for a conventional orbifold interpretation.

\section{Extensions}
\setcounter{equation}{0}

In this section we consider the ``Feigin-Fuchs'' and the inner-automorphic deformations of the general Virasoro construction on orbifold affine algebra.  For $\lambda\ge 3$, further deformations by the full antisymmetric part of the current bilinears $:\hat{J}_{[a}\hat{J}_{b]}:$ may also be possible.
\subsection{$\hat{c}$-changing deformations}
The ``Feigin-Fuchs'' extension is
\alpheqn
\begin{equation}
\hat{T}(z)=\sum_{r=0}^{\lambda-1}\lr_r^{ab}:\hat{J}_a^{(r)}(z)\hat{J}_b^{(-r)}(z):+D^a\partial\hat{J}_a^{(0)}(z)
\end{equation}
\begin{equation}
\lr_r^{ab} =  2 \lr_r^{ac}  \hat{G}_{cd} \lr_r^{db} - \sum^{\lambda-1}_{s=0}\lr_s^{cd} [ \lr_{r+s}^{ef} f_{ce}^{\ \ a}f_{df}^{\ \ b} + f_{ce}^{\ \ f}f_{df}^{\ \ (a} \lr_r^{b) e}]+if_{cd}^{\ \ (a}\lr_r^{b)c}D^d \nonumber \label{FFOVME} 
\end{equation}
\begin{equation}
 0\leq r\leq\hl 
\end{equation}
\begin{equation}
\lr_r^{ab}=\lr_{r \pm \lambda}^{ab}=\lr_{\lambda \pm r}^{ab}, \hspace{.3in}\hl<r\leq\lambda 
\end{equation}
\begin{equation}
D^aM(\lr)_a^{\ b}=D^b, \hspace{.3in}\hat{c}=2 \hat{G}_{ab} (\sum_{r=0}^{\lambda-1}\lr_r^{ab}-6D^aD^b)
\end{equation}
\reseteqn
where the matrix $M(\lr)_a^{\ b}$ is given in (\ref{MN}).  The ground state conformal weight $\hat{\Delta}_0$ is still given by (\ref{delta}), and these generalized $\hat{\textup{c}}$-changing deformations include the familiar $c$-changing deformations$^{\rf{26},\rf{9}}$ at $\lambda=1$.

\subsection{$\hat{c}$-fixed deformations}

Similarly, the generalized  $\hat{\textup{c}}$-fixed deformations have the form 
\alpheqn
\begin{equation}
\hat{T}(z)=\sum_{r=0}^{\lambda-1}\lr_r^{ab}:\hat{J}_a^{(r)}(z)\hat{J}_b^{(-r)}(z):+d^a \frac{\hat{J}_a^{(0)}(z)}{z} + \frac{1}{2z^2}\hat{G}_{ab}d^ad^b
\end{equation}
\begin{equation}
d^aM(\lr)_a^{\ b}=d^b
\end{equation}
\begin{equation}
\hat{c}=2 \hat{G}_{ab} \sum_{r=0}^{\lambda-1}\lr_r^{ab}, \hspace{.3in} \hat{\Delta}_0 = \frac{\hat{G}_{ab}}{2}(\sum_{r=0}^{\lambda -1} \lr_r^{ab} \frac{r (\lambda - r)}{\lambda^2}+d^a d^b)
\end{equation}
\reseteqn
and the OVME (\ref{OVMEgrp}) holds for $\lr_r^{ab}$.  These deformations describe inner automorphic twists or spectral flow in the twisted sectors (see also Subsec. \hspace{-.1in} \ref{doubtwist}), a phenomenon which is familiar$^{\rf{6},\rf{26},\rf{9}}$ at $\lambda=1$.

\subsection{Deformation of the Lie $g$-invariant constructions}

The Lie $g$-invariant constructions of Secs. $5$ and $6$ allow a large class of both the $\hat{c}$-changing and $\hat{c}$-fixed deformations.  In this case we find that 
\alpheqn
\begin{equation}
f_{cd}^{\ \ (a}\lr_r^{b)c} = 0 \label{FFdef}
\end{equation}
\begin{equation}
M(\lr)_a^{\ b}=\Delta \delta_a^{\ b} \ \rightarrow \ D^a(\Delta -1)=0 \label{FFeigen}
\end{equation}
\begin{equation}
\Delta^2=\Delta\ \rightarrow \ \Delta\in\{0,1\} \label{FFeigenc}
\end{equation}
\reseteqn
where $\Delta$ is given in (\ref{DEL}).  The result (\ref{FFdef}) tells us that the extra term in the generalized OVME ({\ref{FFOVME}) vanishes, so that the reduced OVME (\ref{g}) is maintained for $\hat{c}$-fixed or $\hat{c}$-changing deformations of any Lie $g$-invariant construction.

We have recalled in (\ref{FFeigenc}) that all currents $\hat{J}_a^{(0)}$ of any Lie $g$-invariant construction are either (1,0) or (0,0) operators.  Then (\ref{FFeigen}) tells us that one may deform any Lie $g$-invariant construction by arbitrary $D_a$ or $d_a$ for any $\hat{J}_a^{(0)}$ which is a (1,0) current (global $g$ symmetry) of the construction.  In particular the orbifold affine-Sugawara construction may be deformed by arbitrary $D$ or $d$, a fact which is familiar$^{\rf{26}}$ for the affine-Sugawara constructions at $\lambda=1$.  

\subsection{The doubly-twisted affine algebra}
\label{doubtwist}
The $\hat{c}$-fixed deformation of the orbifold affine-Sugawara construction for simple $g$
\namegroup{sflow}
\alpheqn
\begin{equation}
\hat{T}_{g_\lambda}(z;d)=\hat{T}_{g_\lambda}(z)+d^a \frac{\hat{J}_a^{(0)}(z)}{z} + \frac{1}{2z^2}\hat{k}\eta_{ab}d^ad^b,\ \ \forall d
\end{equation}
\begin{equation}
\hat{c}_{g_\lambda}(d)=\lambda c_{g_\lambda}(x),\hspace{.2in}\hat{\Delta}_0^{g_\lambda}(d)=\frac{\hat{c}_{g_\lambda}}{24}(1-\frac{1}{\lambda^2})+\frac{1}{2}\hat{k}\eta_{ab}d^ad^b
\end{equation}
\reseteqn
(see Eqs. (\ref{oassemi}) and (\ref{oassimple})) describes spectral flow in the twisted sectors of the WZW permutation orbifolds
\begin{equation}
\frac{g(d)\stackrel{\lambda\hspace{.05in}times}{\times...\times}g(d)}{{\Z}_{\lambda}}
\end{equation}
where each copy of $g$ has been twisted by the same arbitrary vector $d$, using the known $\lambda=1$ form$^{\rf{26}}$ of (\ref{sflow}).

Following Ref.~\rf{26}, the spectral flow (\ref{sflow}) is equivalent to introducing inner-automorphically twisted orbifold currents.  In the Cartan-Weyl basis, these currents satisfy the ``doubly-twisted'' affine algebra
\namegroup{dta}
\alpheqn
\begin{equation}
[\hat{H}_A^{(r)}(m+\frac{r}{\lambda};d),\hat{H}_B^{(s)}(n+\frac{s}{\lambda};d)]
=\hat{k}(m+\frac{r}{\lambda})\eta_{AB}\delta_{m+n+\frac{r+s}{\lambda},0}
\end{equation}
\begin{equation}
[\hat{H}_A^{(r)}(m+\frac{r}{\lambda};d),\hat{E}_\alpha^{(s)}(n+\frac{s-d\cdot\alpha}{\lambda};d)]
=\alpha_A\hat{E}_\alpha^{(r+s)}(m+n+\frac{r+s-d\cdot\alpha}{\lambda};d)
\end{equation}
\begin{equation}
[\hat{E}_\alpha^{(r)}(m+\frac{r-d\cdot\alpha}{\lambda};d),\hat{E}_\beta^{(s)}(n+\frac{s-d\cdot\beta}{\lambda};d)]\hspace{2.5in}
\end{equation}
\[\hspace{.8in}
=\left\{\begin{array}{lll}
N(\alpha,\beta)\hat{E}_{\alpha+\beta}^{(r+s)}(m+n+\frac{r+s-d\cdot(\alpha+\beta)}{\lambda};d) && \textrm{if}\  (\alpha+\beta)\in\Delta\\
\alpha\cdot\hat{H}^{(r+s)}(m+n+\frac{r+s}{\lambda};d)+\hat{k}(m+\frac{r-d\cdot\alpha}{\lambda})\delta_{m+n+\frac{r+s}{\lambda},0} && \textrm{if}\ (\alpha+\beta)=0\\
0 && \textrm{otherwise}\\
\end{array}\right.
\]
\(
[\hat{H}_A^{(r)}(m+\frac{r}{\lambda};d)-\delta_{m+\frac{r}{\lambda},0}kd_A]|0\rangle=\hat{E}_\alpha^{(r)}(m+\frac{r-d\cdot\alpha}{\lambda};d)|0\rangle \nonumber
=0 {\textrm{\ \ when\ \ }}m+\frac{r}{\lambda}\geq0
\)
\begin{equation}
A,B=1,...,{\textrm{rank}}g,\hspace{.4in}\alpha,\beta\in\Delta
\end{equation}
\begin{equation}
m,n\in\Z,\hspace{.3in}r,s=0,...,\lambda-1,\hspace{.3in}\lambda\in\Z^+
\end{equation}
\reseteqn
for all $d_A$, where $\hat{H}_A$ and $\hat{E}_\alpha$ are the Cartan and root operators respectively.  The relations (\ref{dta}) are understood with the periodicity conditions
\alpheqn
\begin{equation}
\hat{H}_A^{(r\pm\lambda)}(m+\frac{r\pm\lambda}{\lambda};d)=\hat{H}_A^{(r)}(m\pm 1+\frac{r}{\lambda};d)
\end{equation}
\begin{equation}
\hat{E}_\alpha^{(r\pm\lambda)}(m+\frac{r\pm\lambda-d\cdot\alpha}{\lambda};d)=\hat{E}_\alpha^{(r)}(m\pm1+\frac{r-d\cdot\alpha}{\lambda};d)
\end{equation}
\reseteqn
in analogy to the conditions (\ref{percond}).

The doubly-twisted algebra (\ref{dta}) reduces to the orbifold affine algebra (\ref{jalg}) when $d=0$ and it reduces to inner-automorphically twisted$^{\rf{32},\rf{26},\rf{31}}$ affine Lie algebra when $\lambda=1$.  More generally, the doubly-twisted algebra exhibits an intricate interplay between the discrete orbifold twist (outer automorphism) and the continuous spectral flow (inner automorphism).

The doubly-twisted algebra can be obtained by the orbifold induction procedure$^{\rf{1}}$ from the spectral flow at $\lambda=1$ given in Ref.~\rf{26}.  The relations are
\alpheqn
\begin{eqnarray}
\hat{H}_A^{(r)}(m+\frac{r}{\lambda};d)
&\equiv&\hat{J}_A^{(r)}(m+\frac{r}{\lambda})+kd_A\delta_{m+\frac{r}{\lambda},0}\nonumber\\
&=&J_A(\lambda m+r;d)_{BOOST}\nonumber\\
&=&J_A(\lambda m+r)+kd_A\delta_{\lambda m+r,0}
\end{eqnarray}
\begin{eqnarray}
\hat{E}_\alpha^{(r)}(m+\frac{r-d\cdot\alpha}{\lambda};d)
&\equiv&\hat{E}_\alpha^{(r)}(m+\frac{r}{\lambda})\nonumber\\
&=&E_\alpha(\lambda m+r-d\cdot\alpha;d)_{BOOST}\nonumber\\
&=&E_\alpha(\lambda m+r)
\end{eqnarray}
\reseteqn
where the currents $J_A(m;d)_{BOOST},\ E_\alpha(m-d\cdot\alpha;d)_{BOOST}$ are the inner automorphically-twisted currents of Ref.~\rf{26} and $J_A(m),\ E_\alpha(m)$ satisfy affine Lie algebra.
\bigskip

\noindent
{\bf Acknowledgements} 

We thank L. Loveridge and C. Schweigert for helpful discussions.  J.~E and M.~B.~H. were supported in part by the Director, Office of Energy Research, Office of Basic Energy Sciences, of the U.S. Department of Energy under Contract DE-AC03-76F00098 and in part by the National Science Foundation under grant PHY95-14797.

\appendix
\renewcommand{\theequation}{A.\arabic{equation}}
\renewcommand{\thesubsection}{Appendix \Alph{subsection}.}
\subsection{Derivations}
\setcounter{equation}{0}
Many of the operators below have branch points at zero and infinity; however we have checked that all the standard machinery of Ref.~\rf{3} follows using contours that never encircle these points.  \\
\\$\bullet${Wick's theorem}\\
Given the OPE of two operators $A$ and $B$,
\begin{equation}
A(z)B(w)=\sum_{m} \frac{\{AB\}_m(w)}{(z-w)^m}
\end{equation}
the contraction of the two operators is defined as the singular terms of the OPE
\begin{equation}
\contr{.33}{A(z)B(w)}\equiv \sum_{m>0} \frac{\{AB\}_m(w)}{(z-w)^m}.
\end{equation}
For any three operators $A,B,C$ we have
\begin{equation}
\contr{.45}{A(z) :BC:} (w) = \oint_w \frac{(dx)}{x-w} \ \contr{.33}{A(z)B(x)}C(w) + B(x)\contr{.33}{A(z)C(w)}.
\end{equation}
\\$\bullet${Symmetry of the bilinears}\\
To prove the symmetry (\ref{keyident}) of the current bilinears, we use the fact that the commutator of the normal-ordered bilinear $:\hat{J}\hat{J}:$ is related to the derivatives of the singular terms of the $\hat{J}\hat{J}$ OPE (\ref{JJOPE}):
\begin{eqnarray}
:\hat{J}_{(a}^{(r)}\hat{J}_{b)}^{(s)}:(z)-:\hat{J}_{(a}^{(s)}\hat{J}_{b)}^{(r)}:(z) &=& \sum_{m>0} \frac{(-1)^{m+1}}{m!} \partial^m \{\hat{J}_{(a}^{(r)}\hat{J}_{b)}^{(s)}\}_m (z) \nonumber\\
&=&\partial (i f_{(ab)}^{ \ \ c}\hat{J}_{c}^{(r+s)}(z)) + \frac{-1}{2} \partial^2 (\hat{G}_{ab} \delta_{r+s,0\hspace{.05in}mod\hspace{.04in}\lambda}) \nonumber \\
&=&0.
\end{eqnarray}
In this paper, $A_{(a}B_{b)}=A_aB_b+A_bB_a$ and $A_{[a}B_{b]}=A_aB_b-A_bB_a$.\\
\\$\bullet${$\hat{J}:\hat{J}\hat{J}:$ OPE}\\
Starting with the $\hat{J} \hat{J}$ OPE (\ref{JJOPE}) and using Wick's theorem, we can evaluate the $\hat{J}:\hat{J}\hat{J}:$ OPE: 
\aalpheqn
\[
\contr{.75}{\hat{J}_a^{(r)}(z) : \hat{J}_b^{(s)} \hat{J}_c^{(t)}:(w)}= \oint_w \frac{(dx)}{x-w} [\contr{.43}{\hat{J}_a^{(r)}(z) \hat{J}_b^{(s)}(x)} \hat{J}_c^{(t)}(w) + \hat{J}_b^{(s)}(x) \contr{.43}{\hat{J}_a^{(r)}(z) \hat{J}_c^{(t)}(w)}]\nonumber\\
\]
\begin{eqnarray}\hspace{.5in}
& = & \oint_w \frac{(dx)}{x-w} \ (\frac{\hat{G}_{ab}\delta_{r+s,0\hspace{.05in}mod\hspace{.04in}\lambda}}{(z-x)^2} +\frac{if_{ab}^{ \ \ d}\hat{J}_d^{(r+s)}(x)}{(z-x)}) \hat{J}_c^{(t)}(w)\nonumber\\
& \  & + \hat{J}_b^{(s)}(x) (\frac{\hat{G}_{ac}\delta_{r+t,0\hspace{.05in}mod\hspace{.04in}\lambda}}{(z-w)^2} +\frac{if_{ac}^{ \ \ d}\hat{J}_d^{(r+t)}(w)}{(z-w)})\nonumber\\
& = & \frac{i\hat{G}_{dc}f_{ab}^{\ \ d} \delta_{r+s+t,0\hspace{.05in}mod\hspace{.04in}\lambda}}{(z-w)^3}\nonumber\\
& \ & + \frac{(\hat{G}_{ab} \delta_c^e \delta_{r+s,0\hspace{.05in}mod\hspace{.04in}\lambda} + \hat{G}_{ac} \delta_b^e \delta_{r+t,0\hspace{.05in}mod\hspace{.04in}\lambda} -f_{ab}^{\ \ d}f_{dc}^{\ \ e}) \hat{J}_e^{(r+s+t)}(w)}{(z-w)^2}\nonumber\\& \ &  + \frac{i f_{ab}^{\ \ e} :\hat{J}_e^{(r+s)} \hat{J}_c^{(t)}:(w)
+ i f_{ac}^{\ \ d} :\hat{J}_b^{(s)} \hat{J}_d^{(r+t)}:(w)}{z-w}\nonumber\\
&=& \frac{i\hat{G}_{ec}f_{ab}^{\ \ e} \delta_{r+s+t,0\hspace{.05in}mod\hspace{.04in}\lambda}}{(z-w)^3} + \frac{M_{bca}^{(s,t,r) \ e} \hat{J}_e^{(r+s+t)}(w)}{(z-w)^2}\nonumber\\
& \ &  + \frac{i f_{ab}^{\ \ e} :\hat{J}_e^{(r+s)} \hat{J}_c^{(t)}:(w)
+ i f_{ac}^{\ \ e} :\hat{J}_b^{(s)} \hat{J}_e^{(r+t)}:(w)}{z-w}\nonumber\\ \label{J:JJ:OPE}
\end{eqnarray}
\begin{equation}
M_{bca}^{(s,t,r) \ d} \equiv \hat{G}_{ab} \delta_c^d \delta_{r+s,0\hspace{.05in}mod\hspace{.04in}\lambda} + \hat{G}_{ac} \delta_b^d \delta_{r+t,0\hspace{.05in}mod\hspace{.04in}\lambda} -f_{ab}^{\ \ e}f_{ec}^{\ \ d}.
\end{equation}
\areseteqn
\\$\bullet$ ${:\hat{J}\hat{J}:\hat{J} \  OPE}$\\
The $:\hat{J}\hat{J}:\hat{J}$ OPE is obtained by analytic continuation of (\ref{J:JJ:OPE})
\begin{eqnarray}
\contrb{.32}{.7}{:\hat{J}_a^{(r)} \hat{J}_b^{(s)}:(z) \hat{J}_c^{(t)}(w)} &=&  \frac{-i\hat{G}_{ec}f_{ab}^{\ \ e} \delta_{r+s+t,0\hspace{.05in}mod\hspace{.04in}\lambda}}{(z-w)^3}\nonumber\\
& \ & +\frac{M_{abc}^{(r,s,t) \ e} \hat{J}_e^{(r+s+t)}(w)}{(z-w)^2}  + \frac{M_{abc}^{(r,s,t) \ e} \partial_w \hat{J}_e^{(r+s+t)}(w)}{(z-w)}\nonumber\\
& \ &  - (\frac{i f_{ca}^{\ \ e} :\hat{J}_e^{(r+t)} \hat{J}_b^{(s)}:(w) + i f_{cb}^{\ \ e} :\hat{J}_a^{(r)} \hat{J}_e^{(s+t)}:(w)}{z-w}). \label{:JJ:JOPE}
\end{eqnarray}
\\$\bullet${$\hat{T}\hat{J^{(0)}}$} \ OPE\\
Taking $s=-r$ and $t=0$ and multiplying (\ref{:JJ:JOPE}) by $\sum_{r=0}^{\lambda-1} \lr_r^{ab}$, we obtain
\aalpheqn
\begin{equation}
\contr{.4}{\hat{T}(z)\hat{J}_a^{(0)}}(w)= M(\lr)_a^{\ b}(\frac{1}{(z-w)^2}+\frac{\partial_w}{z-w})\hat{J}_b^{(0)}(w)+
\frac{\sum_{r=0}^{\lambda-1} N^{(r)}(\lr)_a^{\ bc} :\hat{J}_b^{(r)}\hat{J}_c^{(-r)}:(w)}{z-w}
\end{equation}
\begin{equation}
M(\lr)_a^{\ b} \equiv \lr_0^{cd} M_{cd,a}^{\hspace{.18in}b} -\sum_{r=1}^{\lambda-1} \lr_r^{cd} f_{ac}^{\ \ f}f_{fd}^{\ \ b} \hspace{.3in} N^{(r)}(\lr)_a^{\ bc} \equiv \lr_r^{de} N_{de,a}^{\hspace{.18in}bc} \label{MN}
\end{equation}
\begin{equation}
M_{ab,c}^{\hspace{.18in}d}\equiv {\delta_{(a}}^d\hat{G}_{b)c}+\frac{1}{2}f_{e(a}^{\ \ \ d}f_{b)c}^{\ \ \ e} \hspace{.3in} N_{ab,c}^{\hspace{.18in}de}\equiv\frac{i}{2}\delta_{(a}^{(d}f_{b)c}^{\ \ \ e)}.
\end{equation}
\areseteqn
\\$\bullet${$\hat{:JJ:}\hat{:JJ:}$ \ OPE}\\
To evaluate the OPE of two current bilinears we apply Wick's theorem again
\aalpheqn
\begin{eqnarray}
\contrb{.32}{1}{:\hat{J}_a^{(r)} \hat{J}_b^{(s)}:(z) :\hat{J}_c^{(t)} \hat{J}_d^{(u)}:(w)} \hspace{-1.3in} && \nonumber \\ 
\hspace{1in}&=& \oint_w \frac{(dx)}{x-w} \contrb{.32}{.7}{:\hat{J}_a^{(r)} \hat{J}_b^{(s)}:(z) \hat{J}_c^{(t)}(x)}\hat{J}_d^{(u)}(w) + \hat{J}_c^{(t)}(x) \contrb{.32}{.7}{:\hat{J}_a^{(r)} \hat{J}_b^{(s)}:(z) \hat{J}_d^{(u)}(w)}\nonumber\\
& = & \oint_w \frac{(dx)}{x-w} [\frac{-i\hat{G}_{ec}f_{ab}^{\ \ e} \delta_{r+s+t,0\hspace{.05in}mod\hspace{.04in}\lambda}}{(z-x)^3}\nonumber\\
& \ & +\frac{M_{abc}^{(r,s,t) \ e} \hat{J}_e^{(r+s+t)}(x)}{(z-x)^2}  + \frac{M_{abc}^{(r,s,t) \ e} \partial_x \hat{J}_e^{(r+s+t)}(x)}{(z-x)}\nonumber\\
& \ &  - (\frac{i f_{ca}^{\ \ e} :\hat{J}_e^{(r+t)} \hat{J}_b^{(s)}:(x) + i f_{cb}^{\ \ e} :\hat{J}_a^{(r)} \hat{J}_e^{(s+t)}:(x)}{z-x})] \hat{J}_d^{(u)}(w)\nonumber\\
& \ & + \hat{J}_c^{(t)}(x) [\frac{-i\hat{G}_{ed}f_{ab}^{\ \ e} \delta_{r+s+u,0\hspace{.05in}mod\hspace{.04in}\lambda}}{(z-w)^3}\nonumber\\
& \ & +\frac{M_{abd}^{(r,s,u) \ e} \hat{J}_e^{(r+s+u)}(w)}{(z-w)^2}  + \frac{M_{abd}^{(r,s,u) \ e} \partial_w \hat{J}_e^{(r+s+u)}(w)}{(z-w)}\nonumber\\
& \ &  \ - (\frac{i f_{da}^{\ \ e} :\hat{J}_e^{(r+u)} \hat{J}_b^{(s)}:(w) + i f_{db}^{\ \ e} :\hat{J}_a^{(r)} \hat{J}_e^{(s+u)}:(w)}{z-w})]
\end{eqnarray}
\begin{eqnarray}
&=& \frac{\hat{G}_{ed} M_{abc}^{(r,s,t) \ e} \delta_{r+s+t+u,0\hspace{.05in}mod\hspace{.04in}\lambda}}{(z-w)^4}\nonumber\\ & & +\frac{-i\hat{G}_{ec}f_{ab}^{\ \ e} \delta_{r+s+t,0\hspace{.05in}mod\hspace{.04in}\lambda}\hat{J}_d^{(u)}(w)-i\hat{G}_{ed}f_{ab}^{\ \ e} \delta_{r+s+u,0\hspace{.05in}mod\hspace{.04in}\lambda}\hat{J}_c^{(t)}(w)}{(z-w)^3}\nonumber\\ 
& \ & + \frac{(if_{fd}^{\ \ e}M_{abc}^{(r,s,t)\ f}-if_{ca}^{\ \ e}M_{ebd}^{(r+t,s,u) \ f} -i f_{cb}^{\ \ e} M_{aed}^{(r,s+t,u) \ f}) \hat{J}_f^{(r+s+t+u)}(w)}{(z-w)^3}\nonumber\\
& & +\{ M_{abc}^{(r,s,t) \ e}:\hat{J}_e^{(r+s+t)} \hat{J}_d^{(u)}:(w) + M_{abd}^{(r,s,u) \ e}:\hat{J}_c^{(t)} \hat{J}_e^{(r+s+u)}:(w)\nonumber\\
& &\ \ \ - f_{ca}^{\ \ e}(f_{de}^{\ \ f}:\hat{J}_f^{(r+t+u)} \hat{J}_b^{(s)}:(w) + f_{db}^{\ \ f}:\hat{J}_e^{(r+t)} \hat{J}_f^{(s+u)}:(w))\nonumber\\
& &\ \ \ -f_{cb}^{\ \ e} (f_{da}^{\ \ f}:\hat{J}_f^{(r+u)} \hat{J}_e^{(s+t)}:(w) + f_{de}^{\ \ f}:\hat{J}_a^{(r)}\hat{J}_f^{(s+t+u)}:(w))\} /(z-w)^2\nonumber\\
& &+ O((z-w)^{-1}). \label{:JJ::JJ:OPE}
\end{eqnarray}
\areseteqn
\vspace{-.35in}
\\$\bullet${Matching to the Virasoro algebra}\\
We turn now to study the general stress tensor $\hat{T}$ in (\ref{stresst}).  To evaluate the $\hat{T}\hat{T}$ OPE set $s=-r$ and $u=-t$ in (\ref{:JJ::JJ:OPE}) and multiply by $\displaystyle{\sum_{r=0}^{\lambda-1} \sum_{t=0}^{\lambda-1} \lr_r^{ab} \lr_t^{cd}}$.  Comparing this to the Virasoro algebra (\ref{viralg}) we find that the third-order pole term is zero since the identities
\aalpheqn
\begin{equation}
\sum_{r,s=0}^{\lambda-1}\lr_r^{ab} \lr_s^{cd} M_{aed}^{(r,-r+s,-s) \ f} f_{cb}^{\ \ e}=0.
\end{equation}
\begin{equation}
\lr_r^{ab} \lr_s^{cd} f_{fd}^{\ \ e} M_{abc}^{(r,-r,s) \ f} 
=\lr_r^{ab} \lr_s^{cd}  f_{ca}^{\ \ e}M_{ebd}^{(r+s,-r,-s) \ f}
\end{equation}
\areseteqn
follow by $a\leftrightarrow b$, $c\leftrightarrow d$ and $ab\leftrightarrow cd$ symmetry of $\sum_{r,s=0}^{\lambda-1}\lr_r^{ab} \lr_s^{cd}$.

Using the symmetry (\ref{keyident}) the restriction (\ref{restric}) is obtained by matching the second-order pole terms.  The fourth-order pole term gives the quadratic form of the central charge
\begin{equation}
\hat{c} = 2 \sum_{r,s=0}^{\lambda-1} \lr_r^{ab} \lr_s^{cd} \hat{G}_{ed} M_{abc}^{(r,-r,s) \ e} 
\end{equation}
which can be simplified to the linear form in (\ref{cc}) using (\ref{restric}).\\
\\$\bullet${First-order pole}\\
The first-order pole term in the Virasoro algebra (\ref{viralg}) is guaranteed to be correct because the higher-order pole terms are correct.  To see this explicitly, we write
\begin{equation}
\hat{T}(z)\hat{T}(w) = \frac{\hat{c}/2}{(z-w)^4} + \frac{2 \hat{T}(w)}{(z-w)^2} + \frac{F(w)}{z-w} + O((z-w)^0) \label{eq:TT1}
\end{equation}
where $F(w)$ is unknown, and this may be relabeled as
\begin{equation}
\hat{T}(w)\hat{T}(z)=\frac{\hat{c}/2}{(w-z)^4} + \frac{2 \hat{T}(z)}{(w-z)^2} + \frac{F(z)}{w-z}+O((z-w)^0) \hspace{.3in}. \label{eq:TT2}
\end{equation}
We can determine $F$ by a Taylor expansion $z=w+(z-w)$ of (\ref{eq:TT1}) and its analytic continuation $\hat{T}(z)\hat{T}(w)=\hat{T}(w)\hat{T}(z)$:
\begin{equation}
\hat{T}(w)\hat{T}(z) = \frac{\hat{c}/2}{(z-w)^4} + \frac{2 \hat{T}(z)}{(z-w)^2} + \frac{2 \partial_z \hat{T}(z)- F(z)}{w-z} + O((z-w)^0) \hspace{.3in}.
\end{equation}
Comparing to (\ref{eq:TT2}) we see that 
\begin{equation}F(w)=\partial_w \hat{T}(w)
\end{equation}
which completes the verification of the Virasoro algebra (\ref{viralg}).\\
\\$\bullet${K-conjugation}\\
In parallel with the VME, it is straightforward to check that the K-conjugate partner $\tilde{\lr}_r^{ab}=(\lr_r^{ab})_{g_\lambda}-\lr_r^{ab}$ of any solution $\lr_r^{ab}$ of the OVME is also a solution of the OVME (use the fact that both $(\lr_r^{ab})_{g_\lambda}$ and $\lr_r^{ab}$ are solutions).
 
To see that the stress tensors of K-conjugate constructions commute, we use (\ref{:JJ::JJ:OPE}) to verify the OPE
\begin{eqnarray}
\hat{T}_{g_{\lambda}}(z)\hat{T}(w)&=&\frac{\hat{c}/2}{(z-w)^4}+\frac{2\hat{T}(w)}{(z-w)^2}+\frac{\partial_w\hat{T}(w)}{(z-w)}+O((z-w)^0)\nonumber\\ &=&\hat{T}(z)\hat{T}(w) + O((z-w)^0) \label{viralg2} 
\end{eqnarray}
and this is easily rearranged into (\ref{commute}).

\renewcommand{\theequation}{B.\arabic{equation}}
\subsection{Graph Theory Ansatz on $g_\lambda=SO(n)_\lambda$}
\setcounter{equation}{0}
The graph theory ansatz $SO(n)_{diag}$ on $SO(n)_{\lambda=1}$ is familiar$^{\rf{28},\rf{2}}$ from the VME.  Here we find the generalization in the OVME to the ansatz $SO(n)_\lambda^{diag}$ on $SO(n)_\lambda$.

The inverse inertia tensor in this case has the form
\begin{equation}
\psi^2 \lr^{ab}_r=\psi^2 \lr_r^{ij,kl}
=\eta^{ij,kl}\lr_r^{ij}
=\delta^{ik}\delta^{jl}\lr_r^{ij},\hspace{.3in} \label{gta}
a=(i,j), \hspace{.3in} 1\leq i<j\leq n
\end{equation}
in the vector index notation for the Cartesian basis of $SO(n)$.  Substituting into the OVME on $g_\lambda=SO(n)_\lambda$, we find the reduced system
\namegroup{SOOVME}
\balpheqn
\begin{equation}
\psi^2 \hat{T}=\sum_{r=0}^{\lambda-1}\sum_{i<j} \lr_r^{ij} :\hat{J}_{ij}^{(r)}\hat{J}_{ij}^{(-r)}:
\end{equation}
\begin{equation}
\lr_r^{ij}(1-\hat{x}\lr_r^{ij})+\tau_n \sum_{s=0}^{\lambda-1} \sum^n_{l\neq i,j}[\lr_s^{il} \lr_{r+s}^{lj} - \lr_r^{ij} (\lr_s^{il} + \lr_s^{jl})]  =0
\end{equation}
\begin{equation} \label{SOOVMEB}
\lr_r^{ij}=\lr_{\lambda \pm r}^{ij}=\lr_{r \pm \lambda}^{ij}, \hspace{.2in} \hl < r \leq \lambda
\end{equation}
\begin{equation}
\lr_r^{ij}=\lr_r^{ji}, \hspace{.4in} \lr_r^{ii}\equiv 0, \hspace{.4in} 
\tau_n=\left\{\begin{array}{lll}1, && n\neq3 \\ 2, && n=3
\end{array}\right. \label{SOOVMED}
\end{equation}
\begin{equation} 
\hat{c} = \hat{x} \sum_{r=0}^{\lambda-1} \sum_{i<j} \lr_r^{ij}, \hspace{,3in} \hat{\Delta}_0=\hat{x} \sum_{r=0}^{\lambda-1} \sum_{i<j} \lr_r^{ij} \frac{r(\lambda-r)}{4 \lambda^2}
\end{equation}
\breseteqn
which further reduces to $SO(n)_{diag}$ when $\lambda=1$.  The conventions in (\ref{SOOVME}d) follow Ref.~\rf{28}.

The reduced OVME in (\ref{SOOVME}) consists of $(\hl+1){n \choose 2}$ equations and unknowns, so the reduced system contains
\begin{equation}
N(SO(n)_{diag},\lambda)=2^{(\hl+1){n \choose 2}}
\end{equation}
level families for each value of $n$ and $\lambda$.  As in the case of the Lie $g$-invariant constructions, most of these level families will be new, with irrational central charges and (for $\lambda\geq 2$) generically irrational ground state conformal weights $\hat{\Delta}_0$.

These level families can be classified by the graphs $G_n^\lambda$ of order $n$ with $\hl+1$ colors, according to the high-level behavior$^{\rf{28},\rf{2}}$:
\balpheqn
\begin{equation}
\lr_r^{ij}(G_n^\lambda)=\frac{\Theta_r^{ij}(G_n^\lambda)}{\hat{x}}+ O(\hat{x}^{-2}), \hspace{.3in} r=0,...,\hl
\end{equation}
\begin{equation}
\psi^2 \hat{T}(G_n^\lambda)=\sum_{r=0}^{\lambda-1} \sum_{i<j} (\frac{\Theta_r^{ij}(G_n^\lambda)}{\hat{x}} + O(\hat{x}^{-2})) :\hat{J}_{ij}^{(r)}\hat{J}_{ij}^{(-r)}:
\end{equation}
\begin{equation}
\hat{c}(G_n^\lambda)=\sum_{r=0}^{\lambda-1} \sum_{i<j} \Theta_r^{ij}(G_n^\lambda) + O(\hat{x}^{-1})
\label{hikc}\end{equation}
\begin{equation}
\hat{\Delta}_0(G_n^\lambda)=\sum_{r=0}^{\lambda-1} \sum_{i<j} \Theta_r^{ij} \frac{r(\lambda-r)}{4 \lambda^2} + O(\hat{x}^{-1})
\end{equation}
\breseteqn
of each level family, where 
\begin{equation}
\Theta_r^{ij}(G_n^\lambda)\in \{0,1\},\hspace{.3in} 1\leq i<j\leq n,\hspace{.3in} r=0,...,\hl
\end{equation}
 is the adjacency matrix of any colored graph $G_n^\lambda$.  The points $i$ and $j$ of $G_n^\lambda$ are connected by an edge of color $r$ when $\Theta_r^{ij}(G_n^\lambda)=1$, and the collection of edges between any $i$ and $j$ is called a bond. 

The high-level expansion also shows that each level family is unitary for $\hat{x}=\lambda x$, $x\in\Z^+$, at least down to some finite radius of convergence $\hat{x}_0=\lambda x_0$.

In the colored graph theory, the orbifold affine-Sugawara construction $SO(n)_\lambda$ lives on the complete colored graph with $n$ points and $\hl+1$ colors in each bond.  As an example, the graph of the orbifold affine-Sugawara construction $SO(3)_{\lambda=2}$ is shown (with colors 0 and 1) below

\begin{picture}(350,140)(0,0)
\put(205,40){0}
\put(161,50){$\bullet$}
\put(160,50){\line(1,0){100}}
\put(162,55){\line(1,0){96}}
\put(205,60){1}
\put(181,100){0}
\put(207,126){$\bullet$}
\put(160,50){\line(3,5){50}}
\put(165,50){\line(3,5){47}}
\put(193,85){1}
\put(235,100){0}
\put(253,50){$\bullet$}
\put(260,50){\line(-3,5){50}}
\put(255,50){\line(-3,5){47}}
\put(223,85){1}
\put(70,15){Figure 4: The orbifold affine-Sugawara construction $SO(3)_{\lambda=2}$}
\end{picture}
\\
so that $\hat{c}=6+O(\hat{x}^{-1})$ in this case. 

Many other properties of the level families in this ansatz can be seen in the colored graphs, in analogy with the application$^{\rf{28}}$ of ordinary graph theory at $\lambda=1$.  As examples,
\\$\bullet${}Aut(SO(n))-inequivalent level families are in one-to-one correspondence with the (point) unlabelled colored graphs.
\\$\bullet${}Since K-conjugation is through the orbifold affine-Sugawara constructions, K-conjugate level families live on complementary colored graphs.
\\$\bullet${}Self K-conjugate constructions$^{\rf{28},\rf{2}}$, with half the orbifold affine-Sugawara central charge
\begin{equation}
\hat{c}=\frac{\hat{c}_{g_\lambda}}{2}
\end{equation}
live on self-complementary colored graphs.

\end{document}